\documentclass{ieeeaccess}
\usepackage{cite}
\usepackage{amsmath,amssymb,amsfonts}
\usepackage{algorithmic}
\usepackage{graphicx}
\usepackage{textcomp,longtable}
\usepackage{amsmath,amsfonts,supertabular}
\usepackage{algorithmic}
\usepackage{algorithm}
\usepackage{array}
\usepackage[caption=false,font=normalsize,labelfont=sf,textfont=sf]{subfig}
\usepackage{makecell}

\usepackage{textcomp}
\usepackage{stfloats}
\usepackage{url}
\usepackage{verbatim}
\usepackage{graphicx}
\usepackage{cite}
\usepackage{amssymb,color}

\usepackage{bm}
\makeatletter
\AtBeginDocument{\DeclareMathVersion{bold}
\SetSymbolFont{operators}{bold}{T1}{times}{b}{n}
\SetSymbolFont{NewLetters}{bold}{T1}{times}{b}{it}
\SetMathAlphabet{\mathrm}{bold}{T1}{times}{b}{n}
\SetMathAlphabet{\mathit}{bold}{T1}{times}{b}{it}
\SetMathAlphabet{\mathbf}{bold}{T1}{times}{b}{n}
\SetMathAlphabet{\mathtt}{bold}{OT1}{pcr}{b}{n}
\SetSymbolFont{symbols}{bold}{OMS}{cmsy}{b}{n}
\renewcommand\boldmath{\@nomath\boldmath\mathversion{bold}}}
\makeatother

\def\BibTeX{{\rm B\kern-.05em{\sc i\kern-.025em b}\kern-.08em
    T\kern-.1667em\lower.7ex\hbox{E}\kern-.125emX}}

\begin{document}
\history{Date of publication xxxx 00, 0000, date of current version xxxx 00, 0000.}
\doi{10.1109/ACCESS.2024.0429000}

\title{Computing in Integrated Terrestrial and Non-Terrestrial Networks: A Comprehensive Survey}
\author{\uppercase{Hoe Ziet Wong}\authorrefmark{1},
\uppercase{Insaf Rzig}\authorrefmark{2}, \IEEEmembership{Graduate Student, IEEE} \uppercase{Safwan Alfattani}\authorrefmark{3}, \IEEEmembership{Member, IEEE}, and \uppercase{Wael Jaafar} \authorrefmark{4}, \IEEEmembership{Senior Member, IEEE}}

\address[1]{Télécom SudParis, Paris, 91000 France (e-mail: hoe-ziet.wong@telecom-sudparis.eu)}
\address[2]{MEDIATRON Laboratory, École Supérieure des Communications (Sup'Com), University of Carthage, Ariana, 2083 Tunisia (e-mail: insaf.rzig@supcom.tn)}
\address[3]{Department of Software and IT Engineering, École de technologie supérieure (ÉTS), University of Quebec, Montreal, QC H3C 1K3 Canada (e-mail: wael.jaafar@etsmtl.ca)}
\address[4]{King AbdulAziz University, Rabigh, 25732 Saudi Arabia (e-mail: smalfattani@kau.edu.sa)}
\tfootnote{This work is in part funded by the EU Erasmus+ scholarships program and in part by a King AbdulAziz University grant.}

\markboth
{Author \headeretal: Preparation of Papers for IEEE TRANSACTIONS and JOURNALS}
{Author \headeretal: Preparation of Papers for IEEE TRANSACTIONS and JOURNALS}

\corresp{Corresponding author: W. Jaafar (e-mail: wael.jaafar@etsmtl.ca).}

\begin{abstract}
The rapid growth of Internet-of-things (IoT) devices, smart vehicles, and other connected objects is driving demand for ubiquitous connectivity and intensive computing capacity. 5G and upcoming 6G networks are crucial to meeting these demands and the fast-evolving services and applications. However, traditional terrestrial networks face limitations in coverage and capacity. Integrated Terrestrial and Non-Terrestrial Networks (ITNTN) are emerging to address these challenges. In essence, ITNTN combines ground-based infrastructure with aerial, space, and water surface networks to provide seamless connectivity and computing resources anytime, anywhere. Given the stringent quality-of-service (QoS) of future services, edge computing will be an inseparable component of ITNTN. Consequently, we dive in this survey into current efforts of integrating cloud/fog/edge computing into ITNTN layers to facilitate stringent QoS services and address the data processing needs of modern applications. 
Since there have been only limited and partial efforts in integrating computing functionalities within ITNTN, we aim to extend the discussion to the full integration of computing and identifying the challenges and future research directions to achieve it. 
\end{abstract}


\titlepgskip=-21pt

\maketitle

\section{Introduction}
\label{sec:introduction}
\PARstart{T}{he} proliferation of smart connected objects, including Internet of Things (IoT) devices, smart vehicles, mobile phones, and drones, has been accelerating rapidly in recent years. The number of connected devices is projected to reach 29 billion by 2030. This unprecedented growth is driven by the demand for ubiquitous and seamless connectivity. The development of 5G and the upcoming 6G networks have significantly contributed to this trend. Notably, three key visions for 6G are advanced virtual communications, extensive connectivity of devices, and ultra-low latency communications \cite{9390169}. The development of smart cities and 6G networks are closely intertwined. Specifically, 6G will enable a highly connected network, advancing systems such as intelligent transport systems (ITS), power systems, agriculture, and more. Furthermore, low latency services are a top priority for service providers, not only for 5G and 6G networks but also for streaming services, which account for 60-70\% of Internet traffic. Consequently, the rise in the number of devices and the need to support these services have prompted the demand for more computing resources and a network with seamless connectivity. 

To address the aforementioned challenges, a recent paradigm known as Integrated Terrestrial Network and Non-Terrestrial Network (ITNTN)-based computing has been introduced. By leveraging the communication and computing resources available in ITNTN, services with stringent quality-of-service (QoS) requirements can be effectively supported. Typically, ITNTN consists of two main components: terrestrial networks (TN) and non-terrestrial networks (NTN).

As illustrated in Fig. \ref{fig:itntn}, the ground infrastructures of TN consist of base stations (BSs) of cellular networks and roadside units (RSUs). Moreover, the TN may include connected and autonomous vehicles (CAVs) and IoT devices. Currently, TN is still the principal network that ensures communication and connectivity on the ground. Nonetheless, the deployment of BSs is often concentrated in areas with high to medium populations due to economic considerations. Maintaining such infrastructure in sparsely populated regions is often economically unsustainable for service providers, leading to coverage and performance issues for residents in these areas.
Even in urban areas, ensuring network coverage has proven challenging due to obstacles and limited space, which limits the network's scalability. The ground infrastructure is also prone to damage from natural disasters and human activities. As for CAVs and IoT devices, they rely on the ground infrastructure, given their limited computing capacity. Nevertheless, the TN infrastructure might not be available everywhere, and other solutions must be introduced to complement it.

\Figure[t!](topskip=0pt, botskip=0pt, midskip=0pt)[width=1.5\columnwidth]{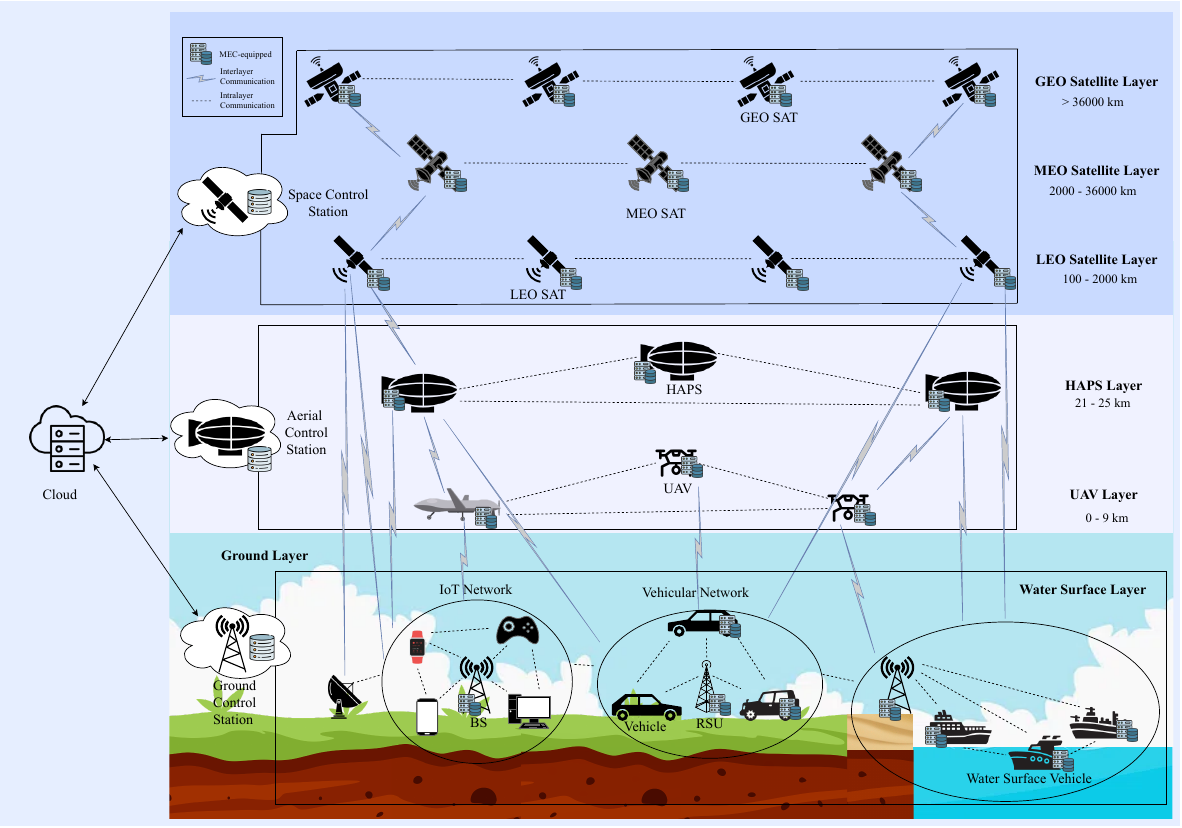}
{ \textbf{The framework of ITNTN-based computing.}\label{fig:itntn}}

One solution relies on deploying a computing-based NTN infrastructure. NTN involves several network components, such as unmanned aerial vehicles (UAVs), high-altitude platform stations (HAPS), satellite networks (STNs), and water-surface networks (WSNs), illustrated in Fig. \ref{fig:itntn}. UAVs, also better known as drones, have been used extensively in different applications, notably in the military for surveillance, intelligence gathering, telecommunications, etc. In public settings, UAVs can serve as aerial BSs (UAV-BSs) and wireless relays, thus improving communications in areas lacking the ground infrastructure or in emergencies \cite{9535463}. A civilian commercial UAV can fly up to a few kilometers from the Earth's surface \cite{9535463}. Higher in the sky, the HAPS can be positioned in the stratosphere at an altitude of 17-25 km, and its coverage zone can vary from 36 km to 234 km (ITU-R F.1500). However, current HAPS projects typically aim for less than 100 km coverage radius to achieve high-area throughput. HAPS can take the form of an aircraft, a glider, an airship, or a balloon. It can be either continuously flying or quasi-stationary. By adapting its trajectory or being steady in the sky, it can provide a stable coverage footprint \cite{9380673}. In contrast, satellite systems are known for their near-global or wide coverage footprint compared to UAV and HAPS. They also have longer operation live spans and are designed to withstand harsh space environments. Thus, they are less vulnerable to weather conditions, making them a reliable communication system. According to ITU-R S.1328-2, there are three types of satellite systems: 1)a typical Low Earth Orbit (LEO) satellite system can be deployed at an altitude ranging between 100 km and 2000 km, 2) Medium Earth Orbit (MEO) satellites have an orbital attitude between 2000 km and 36000 km, while 3) Geostationary Earth Orbit (GEO) systems operate from the altitude of 36000 km. Finally, a WSN is composed of manned/unmanned sea-surface vehicles or ships, which can interconnect through device-to-device (D2D) links or to other NTN equipment such as UAVs, HAPS, or satellites.

Although NTN complements TN to provide global connectivity, it is still challenging to satisfy services with intensive computing requirements or stringent delay requirements that cannot be satisfied locally. To satisfy the demand of computation-intensive applications, cloud computing was initially introduced to allow user devices to offload their tasks to servers in the cloud infrastructure \cite{6553297}. Although cloud computing is widely deployed, it cannot respond to critical services with stringent delay QoS, such as CAVs and public safety services \cite{8100873}. To address this issue, edge computing was introduced, where computing units are deployed in the proximity of end users. Computing power can be embedded within ground BSs, RSUs, CAVs, UAV-BSs, HAPS, STNs, WSNs, etc., to handle computing-intensive tasks. However, the limited energy available at each type of these nodes might influence the way of using them to offload computing tasks.

The open technical literature on computing-based ITNTN has greatly matured, particularly concerning communication and computing in TN \cite{7879258}, in one or several NTN layers \cite{9137644,9535463,9861699, 10089158,10045716,bakambekova2024interplayartificialintelligencespaceairground}, and with or without the TN layer \cite{10213423,9982444,Mahboob_2024}. However, little attention has been given to the fully-integrated ITNTN \cite{8368236,9992172,RAGO2024110725,10745905}. Despite the significant potential of ITNTN, comprehensive literature surveys specifically focusing on ITNTN-based computing have not yet been extensively explored.   

Motivated by the above, this contribution sheds light on the current task offloading techniques in TN and NTN network layers before exploring the real potential of ITNTN-based computing. Specifically, we begin by presenting a comprehensive study of the computing concepts, including cloud, fog, and edge computing. Having this in mind, we overview current TN and NTN-based computing systems. Then, we address the full integration of ITNTN with the computing paradigm. Finally, open issues and some interesting research directions are discussed.
To the best of our knowledge, this is the first comprehensive survey that highlights the importance of computing within the ITNTN framework, including extensive TN and NTN network components. Table \ref{tab:surveys} below summarizes previous surveys on ITNTN, highlighting their scope, key contributions, and the specific ITNTN components addressed in each study.

\begin{table*}[t]
    \centering
    \large
    \caption{Summary of Related Surveys}
    \label{tab:surveys}
    \resizebox{1\textwidth}{!}{
    \begin{tabular}{|c|c|c|c|c|c|c|c|c|c|c|l|l|}
        \hline
        \textbf{Ref.} & \textbf{Year} & \multicolumn{2}{c|}{\textbf{Covered Areas}} & \multicolumn{7}{c|}{\textbf{Network Components}} & \textbf{Contributions} \\
        \cline{3-4} \cline{5-11}
        & & \textbf{Commun.} & \makecell{\textbf{Edge}\\ \textbf{Comput.}} & \makecell{\textbf{Ground} \\ \textbf{Infra.}} & \textbf{Sat.} & \textbf{HAPS} & \textbf{UAV} & \makecell{\textbf{Water}\\ \textbf{Surface Equi.}}  & \makecell{\textbf{Veh.}\\ \textbf{Netw.}} & \makecell{\textbf{IoT} \\ \textbf{Integr.}} & \\
        \hline
        \cite{7879258} & 2017 & Partial & \checkmark & \checkmark &  &  &  &  &  &  & \makecell[l]{$\circ$ Review and classification of computation\\ offloading techniques in MEC. \\ $\circ$ Summary of resource allocation \\techniques in MEC.}
         \\
        \hline
        \cite{8368236} & 2018 & \checkmark &  & \checkmark & \checkmark & \checkmark & \checkmark &  & \checkmark &  & 
        
        \makecell[l]{$\circ$ Frameworks and architectures of ITNTN. \\ Communication models (mobility management, \\ spectrum management, traffic offloading, etc.) \\ Challenges and future research directions}
         \\
        \hline
        \cite{9137644} & 2020 & \checkmark & \checkmark & \checkmark &  & \checkmark & \checkmark &  &  &  & 

        \makecell[l]{$\circ$ UAV deployment, computation offloading,\\ and caching techniques in UAV-enabled MEC.} \\
        \hline
        \cite{9535463} & 2021 & \checkmark & \checkmark & \checkmark &  &  & \checkmark &  & \checkmark &  & 
        \makecell[l]{$\circ$ Review of UAV-MEC architectures. \\ $\circ$ Overview of computation offloading and resource \\allocation techniques in UAV-MEC systems.}
        \\
        \hline
        \cite{9992172} & 2022 & \checkmark &  & \checkmark & \checkmark & \checkmark & \checkmark &  & \checkmark &  & 
        
        \makecell[l]{$\circ$ Use cases of ITNTN. \\ $\circ$ Consideration of two multi-operator paradigms: \\ Cellular Net. + UAVs, and Cellular Net. + STN.}
       \\ 
        \hline
        \cite{9861699} & 2022 & \checkmark & Partial & \checkmark & \checkmark &  & \checkmark &  & \checkmark &  & 
        
        \makecell[l]{$\circ$ Review on the integration of NTN in 5G \\ (Architecture, communication, and computing). \\ $\circ$ Review on ML, mmWave, and THz for NTN. \\ Integration of NTN in 6G.} \\
        \hline
        \cite{9982444} & 2022 & \checkmark & Partial & \checkmark & \checkmark &  &  &  &  &  & 
        \makecell[l]{$\circ$ Redefinition and architecture of space networks. \\ $\circ$ Discussion of space networks' enabling\\ technologies.}
        \\
        \hline
        \cite{10089158} & 2023 & \checkmark & \checkmark & \checkmark & \checkmark &  & \checkmark &  &  &  & 
        
        \makecell[l]{$\circ$ Definition of the aerial edge computing\\ architecture. \\ $\circ$ Studies classification based on achieved\\ performances.}\\
        \hline
        \cite{10045716} & 2023 & \checkmark & \checkmark & \checkmark & \checkmark & \checkmark & \checkmark &  &  &  & 
        
        \makecell[l]{$\circ$ Architectures of NTN networks for 6G. \\ $\circ$ Enabling technologies of NTNs \\ (SDN, AI, computation offloading, etc.). \\ Optimization techniques of communication\\ resources.}
        \\
        \hline
        \cite{10213423} & 2023 & Partial & Partial & \checkmark & \checkmark &  &  &  &  &  & 
        \makecell[l]{$\circ$ Development, visions and challenges \\ of LEO satellite computing. \\ $\circ$ Presentation of Tiansuan constellation\\ research platform.}
        \\
        \hline
        \cite{bakambekova2024interplayartificialintelligencespaceairground} & 2024 & Partial & Partial & \checkmark & \checkmark & \checkmark & \checkmark &  &  &  & 
        \makecell[l]{$\circ$ AI techniques for computation offloading. \\ $\circ$ Resource allocation optimization in UAVs,\\ HAPS, and satellites.}
        \\
        \hline
        \cite{Mahboob_2024} & 2024 & \checkmark & Partial &  & \checkmark &  &  &  &  &  & 
        \makecell[l]{$\circ$ Insights on integrating AI to \\satellite-based NTN. \\ $\circ$ Review of AI-NTN computing.}
        \\
        \hline
        \cite{RAGO2024110725} & 2024 & \checkmark & & \checkmark & \checkmark & \checkmark & \checkmark &  &  &  \checkmark &
        \makecell[l]{$\circ$ Use cases of ITNTN networks. \\  $\circ$ KPIs and requirements of ITNTN.\\ $\circ$ Different ITNTN architectures.}
        \\
        \hline
        \cite{10745905} & 2024 & \checkmark & Partial & \checkmark & \checkmark & \checkmark & \checkmark &  &  &  &
        \makecell[l]{$\circ$ Foundations and principles of ITNTN networks. \\  $\circ$ Enabling technologies of ITNTN.\\ $\circ$ Future challenges in ITNTN development.}
        \\
        \hline
        \makecell{\textbf{Our} \\ \textbf{Survey}} & 2025 & {Partial} & {\checkmark} & {\checkmark} & {\checkmark} & {\checkmark} & {\checkmark} & {\checkmark} & {\checkmark} & {\checkmark} & 
        \makecell[l]{$\circ$ Review of edge computing in separate \\and integrated ITNTN layers. \\ $\circ$ Classification of task offloading and resource\\ management approaches.\\ $\circ$ Identification of ITNTN edge computing issues\\ and future research.}
        \\
        \hline
    \end{tabular}%
    }
\end{table*}

The rest of the paper is organized as follows. Section \ref{sec:fundamentalcomputing} focuses on the fundamental principles of computing, which include the definitions of different types of computing, their characteristics, and examples. Section \ref{sec:tn} overviews computing techniques in terrestrial networks, whereas Section \ref{sec:ntn} is dedicated to NTN-based computing. Subsequently, ITNTN-based computing is discussed in Section \ref{sec:itntn}, while Section \ref{sec:future} identifies the challenges and future research directions. Finally, Section \ref{sec:end} concludes the survey.


The list of abbreviations regularly employed in this article is presented in Table \ref{tab:abbr}.

\begin{footnotesize}
\begin{table*}[!ht]
\centering
\caption{List of Relevant Abbreviations}
    \label{tab:abbr}
   \begin{tabular}{p{1.2cm} p{6.5cm} p{1.2cm} p{6.5cm}}
        AC & Actor-Critic &
        ADP & Approximate Dynamic Programming\\
        AI & Artificial Intelligence &
        AO & Alternating Optimization \\
        AR & Augmented Reality &
        AREA & Application-awaRE workload Allocation \\
        BCD & Block Coordinate Descent &
        BFDA & Breakeven-Free Double Auction \\
        BS & Base Station &
        DDQN & Double Deep Q Network \\
        DECO & Differential Evolution-based Convex Optimization &
        DDPG & Deep Deterministic Policy Gradient \\
        DL & Deep Learning &
        DPCOEM & Dynamic Parallel Computing Offloading and Energy Management \\
        DQN & Deep Q Network &
        DRL & Deep Reinforcement Learning \\
        DT & Digital Twin &
        EH & Energy Harvesting \\
        FL & Federated Learning &
        GA & Genetic Algorithm \\
        GBA & Greedy Bat algorithm &
        GD & Ground Device \\
        GEO & Geostationary Earth Orbit &
        GEOM & Geometric Optimization Method\\
        GES & Greedy-based Efficient Searching &
        HAPS & High Altitude Platforms \\
        IaaS & Infrastructure as a Service &
        IoT & Internet-of-Things \\
        IoV & Internet-of-Vehicles &
        IRS & Intelligent Reflecting Surface \\
        ITNTN & Integrated Terrestrial \& Non-Terrestrial Network &
        ITS & Intelligent Transport System \\
        ITU & International Telecommunication Union &
        JAOBSS & Joint Alternative Optimization-based Bi-Section Searching \\
        JODTS & Joint Offloading Decision and Task Scheduling &        
        JRACO & Joint Resource Allocation and Computation Offloading \\ 
        KKT & Karush-Kuhn-Tucker &
        LDD & Lagrange Dual Decomposition \\
        LEO & Low Earth Orbit &
        LMAPPO & Lyapunov-guided Multi-Agent Proximal Policy Optimization \\
        LLM & Large Language Model &
        LSTM & Long Short-Term Memory \\
        MADRL & Multi-Agent Deep Reinforcement Learning &
        MDP & Markov Decision Process \\
        MEC & Mobile Edge Computing &
        MEO & Medium Earth Orbit \\
        MINLP & Mixed-Integer Nonlinear Programming &
        ML & Machine Learning \\
        MU-MIMO & Multi-User Multiple Input and Multiple Output &
        MR & Mobile Relay \\
        NFV & Network Function Virtualisation &
        NIST & National Institute of Standards and Technology \\
        NSGA-II & Non-dominated Sorting Genetic Algorithm II &
        NGSA-III & Non-dominated Sorting Genetic Algorithm III \\
        NTN & Non-Terrestrial Network &
        OCRA & Optimized Computing Resource Allocation \\
        ORAN & Open Radio Access Network &
        OTAA & Optimized Task Allocation Approach \\
        PER & Prioritized Experience Replay &
        PNE & Pure Nash Equilibrium \\
        PSO & Particle Swarm Optimization &
        QoE & Quality-of-Experience \\
        QoS & Quality-of-Service &
        RAN & Radio Access Network \\
        RL & Reinforcement Learning &
        RSU & Roadside Unit \\
        SaaS & Software as a Service &
        SAC & Soft Actor-Critic \\
        SAGIN & Space-Air-Ground Integrated Network &
        SAT & Satellite \\
        SCA & Successive Convex Approximation &
        SCR & Sum Computation Rate \\
        SDN & Software-Defined Network &
        SDP & Semi-Definite Program \\
        SLA & Service-Level Agreement &
        STN & Satellite Network \\
        SVM & Support Vector Machine &
        TD3 & Twin Delayed Deep Deterministic Policy Gradient \\
        TN & Terrestrial Network &
        UAV & Unmanned Aerial Vehicle \\
        V2I & Vehicle-to-Infrastructure &
        V2V & Vehicle-to-Vehicle \\
        VEC & Vehicular Edge Computing &
        WCPN & Wireless Computing Power Network \\
    \end{tabular}
\end{table*}
\end{footnotesize}

\section{Fundamentals of Computing}\label{sec:fundamentalcomputing}
In this section, we provide a brief overview of the fundamental concepts of cloud, fog, and edge computing, whose architecture is illustrated in Fig. \ref{fig:cfe}.

\Figure[t!](topskip=0pt, botskip=0pt, midskip=0pt)[width=0.99\columnwidth]{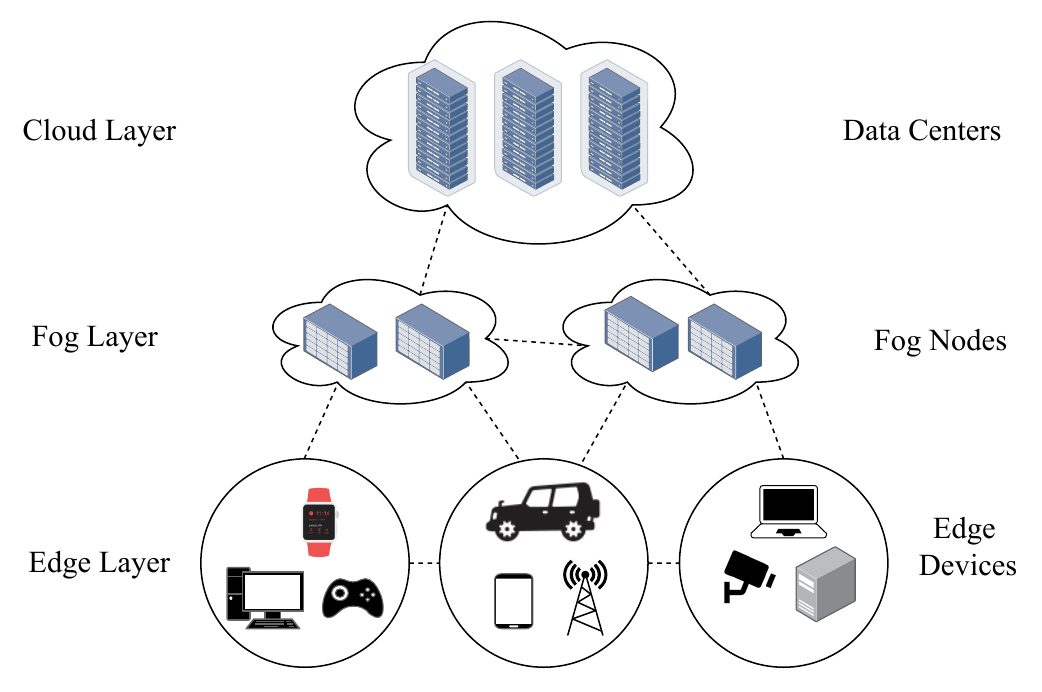}
{ \textbf{Architecture of cloud, fog and edge computing.}\label{fig:cfe}}

\subsection{Cloud Computing}

The National Institute of Standards and Technology (NIST) defines cloud computing as ``a model for enabling ubiquitous, convenient, on-demand network access to a shared pool of configurable computing resources (e.g., networks, servers, storage, applications, and services) that can be rapidly provisioned and released with minimal management effort or service provider interaction'' \cite{14886}. In brief, cloud computing can be regarded as a paradigm that describes a network of omnipresent and shared computing resources, which is used to provide different and personalized services according to service-level agreements (SLAs). 

According to \cite{14886}, cloud computing is an \textit{on-demand self-service}, i.e., computing resources can be accessed at any time by the consumer without the intervention of the provider. It also provides a \textit{broad network access}, i.e., resources are accessible from any type of platform, e.g., mobile phones, laptops, etc. Computing resources support \textit{resource pooling} to handle distribution to different users simultaneously. They are also \textit{elastic}, i.e., they can be allocated and recalled dynamically among consumers and their needs. The use of cloud computing resources is a \textit{measured service} where usage is quantified and monitored, thus providing transparency for the consumers. Finally, cloud computing resources are \textit{available} through redundancy and resilience mechanisms and are \textit{scalable} to adapt to dynamic consumer needs. 

Three types of cloud computing services can be provided, namely Infrastructure as a Service (IaaS), Platform as a Service (PaaS), and Software as a Service (SaaS).
In IaaS, the infrastructure (physical and/or virtual resources) is provided to the consumers in the form of storage, hardware, processing units, etc. IaaS allows consumers to reduce investment in hardware requirements. Digital Ocean, Microsoft Azure, and AWS EC2 are known IaaS platforms.
In PaaS, the development and execution environment is provided as an overlay to the infrastructure. This framework facilitates software development. Such work can be done on Google App Engines, AWS Elastic Beanstalk, Apache Stratos, etc. Finally, SaaS provides the core functionalities to the consumer without regard to the underlying infrastructure or development framework. Examples of SaaS include Salesforce, Dropbox, DocuSign, etc.

Cloud computing platforms can be deployed according to different models. Each model offers different levels of control, flexibility, and management. As shown in Fig. \ref{fig:typecloud}, a cloud computing platform can be deployed as a \textit{Public Cloud} where all resources are shared and accessible to all tenants subscribing to the service. The public cloud has a low cost at the expense of security. Moreover, a cloud can be {private}. A \textit{Private Cloud} is privately owned by a single tenant and is designed with better security and protection than a public cloud.     
Alternatively, the \textit{community Cloud} has been created to allow organizations to collaborate by sharing data and services between themselves. Finally, a \textit{Hybrid Cloud} can be a combination of public and private clouds, thus offering different levels of customization to consumers balancing between scalability and security.

\Figure[t!](topskip=0pt, botskip=0pt, midskip=0pt)[trim={0.5cm 1.8cm 0.6cm 1.5cm},clip,width=0.8\columnwidth]{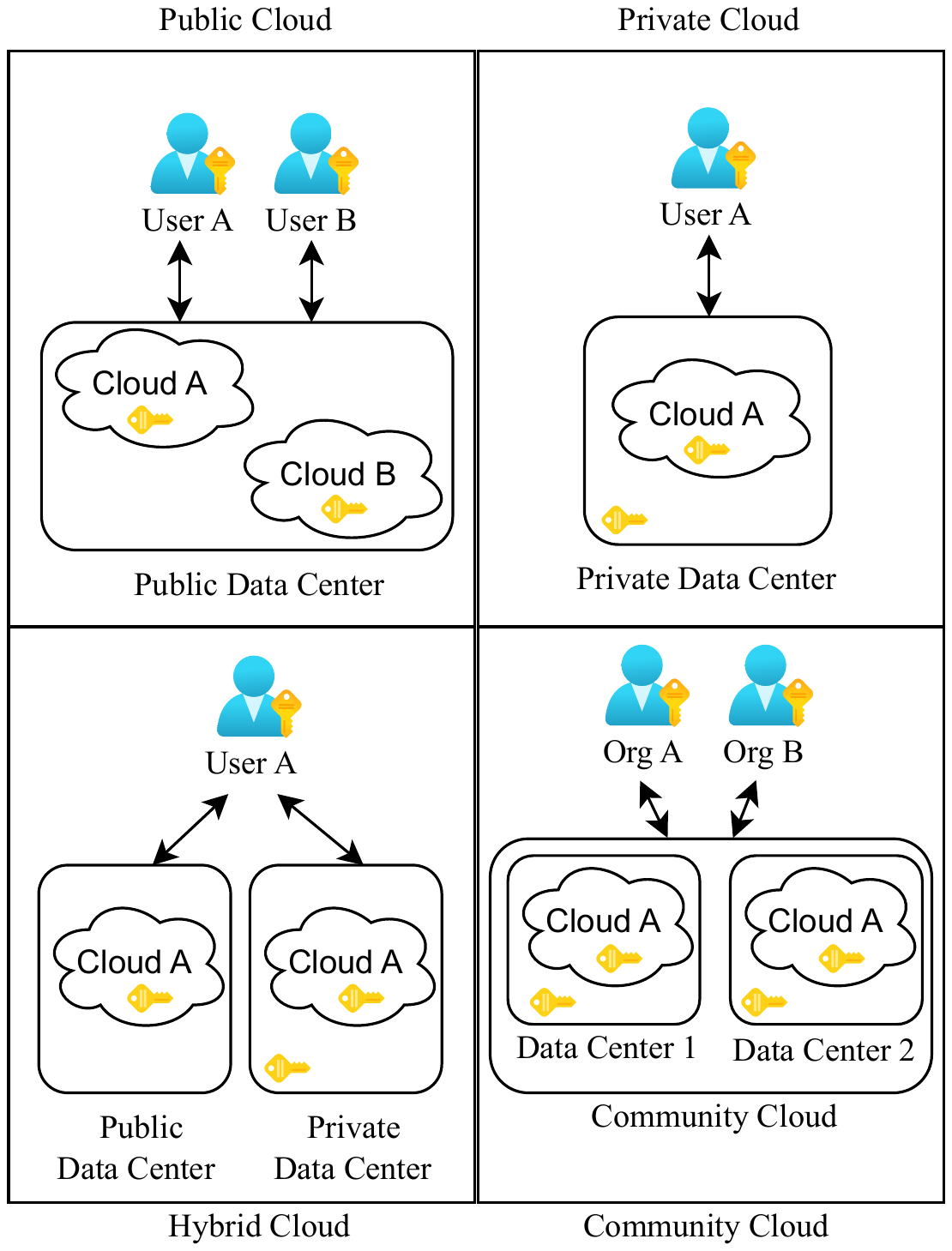}
{ \textbf{Deployment models of cloud computing platforms.}\label{fig:typecloud}}

\subsection{Fog Computing}

The OpenFog Consortium, which was created to accelerate the adoption and advancement of fog computing, defines it as ``a system-level horizontal architecture that distributes resources and services of computing, storage, control and networking anywhere along the continuum from Cloud to Things, thereby accelerating the velocity of decision-making.'' Meanwhile, the definition of Fog Computing by NIST \cite{235721} is given by ``Fog computing is a layered model for enabling ubiquitous access to a shared continuum of scalable computing resources. The model facilitates the deployment of distributed, latency-aware applications and services and consists of fog nodes (physical or virtual), residing between smart-end devices and centralized (cloud) services. The fog nodes are context-aware and support a common data management and communication system.'' In summary, fog computing is a distributed architecture of computing nodes situated anywhere between end devices and cloud networks, not solely at the edge. Fog computing is capable of providing computing, storage, and networking services while assuring low latency, energy cost, and high bandwidth efficiency. However, it is important to note that fog computing is a non-standalone system as it depends on the cloud and its multi-tier architecture, allowing for high flexibility \cite{8100873}.

Fog computing bridges the gap between centralized data centers and the vast array of end devices via its two main characteristics. The first characteristic is the flexible placement of fog nodes, enabling the processing of a significant amount of data with low latency \cite{9435363}. By processing data closer to the edge network, the processing delay and communication time can be reduced. This will eventually save the bandwidth and improve the overall efficiency of the system. The second characteristic is the capacity of fog computing to provide adequate resources to end devices. Indeed, unlike edge computing, which has scarce resources (mainly due to physical and spatial limitations),  the flexible placement of fog computing nodes can subdue this issue since it is not exclusively situated at the edge of the network \cite{8869772}. By sitting between cloud and edge networks, fog networks can be seen as a win-win solution for a compromise between computing resources and latency.

\subsection{Edge Computing}

Cloud computing is often regarded as an example of a centralized computing paradigm, where the infrastructure, such as data centers, are usually located in the same area and are far from end users. Due to distance, critical services such as public safety and autonomous driving cannot be efficiently handled by cloud computing. Alternatively, edge computing has been introduced as a viable solution to process delay-sensitive applications. Typically, edge computing is defined as a distributed computing paradigm that aims to bring computing resources and data storage the closest to the end users in an attempt to improve the response time and save bandwidth. Cao \textit{et al.} defined it in \cite{9083958} as a ``new computing paradigm that performs computing at the edge of the network. Its core idea is to make computing closer to the source of the data.'' 

Edge computing is designed based on the following principles: Its distributed infrastructure is assured to be the \textit{closest to end users}. With such a proximity, it achieves \textit{low-latency} task offloading. Task processing can be \textit{split among several edge nodes} at different locations, thus ensuring \textit{low bandwidth usage and cost}. With its distributed infrastructure, edge computing is highly \textit{scalable}. Finally, through its localized processing, it \textit{reduces the risk} of data breach and exposure \cite{9083958}. 

\section{Computing in Terrestrial Networks}\label{sec:tn}

This section is dedicated to reviewing work on the integration of edge computing in ground systems, including maritime infrastructure, vehicular networks, and IoT networks.

\subsection{ Ground Network-based Computing}

\begin{table*}[t]
    \centering
    \Large
    \caption{Related Work on Computing in Terrestrial Networks}
    \label{tab:ground}
    \resizebox{\textwidth}{!}{%
    \begin{tabular}{|c|c|c|l|l|}
        \hline
        \textbf{Ref} & \textbf{System Model} & \textbf{Objective} & \textbf{Proposed Solutions} & \textbf{Findings} \\
        \hline
        \cite{8493145} & 
        \makecell{Cloud computing + BSs \\+ WiFf APs + Users}
        & \makecell{Min. latency in  \\non-blocking state  \\  + Min. recovery time\\ in blocking state}
        & 
        \makecell[l]{EdgeFlow: MDP-based partial task offloading\\ scheme for local, cloudlet, and cloud computing.}

       & \makecell[l]{EdgeFlow adapts to varying data generation speeds.\\ EdgeFLow achieves $15 \%$ (resp. $43 \%$) higher processing \\rate than MDP scheme (resp. local computing).}
       \\
        \hline
        \cite{8732365} & \makecell{MEC-equipped normal \\and green BSs + Users
        }
        
        & \makecell{Min. total price-weighted \\energy consumption} & 
        \makecell[l]{Mixed timescale optimization + joint\\ optimization of offloading  strategy and computing\\ resource allocation using monotonicity analysis.}
        & \makecell[l]{Total price-weighted energy consumption decreases \\with higher SNR and number of MEC servers.
        \\}
        \\
        \hline
        \cite{9956845} & \makecell{Cloud computing + MEC \\servers + BSs + Users}
        & \makecell{Min. task completion latency \\ + Min. energy consumption}
         & \makecell[l]{WCPN with multi-agent DRL scheme for task\\ scheduling and computing resource allocation. 
         }
          & \makecell[l]{Proposed DRL scheme achieves best latency \\performances compared to benchmarks.}
        \\
        \hline
        \cite{9803864} & \makecell{MEC-equipped BSs \\+ Mobile relays \\+ Train users}
        & \makecell{Min. average \\task execution latency} & 
        \makecell[l]{JRACO framework based on game theory \\to optimize communication and computing\\ resource allocation.} & 
        \makecell[l]{JRACO achieves the best latency performance\\ compared to benchmarks for different parameters, \\e.g., task size, number of users, number of channels,\\ and computation capacity.}
        \\
        \hline
        \cite{10507238} & 
        \makecell{MEC-equipped BSs \\+ Users} & Min. task offloading delay & 
        \makecell[l]{Joint Tammer decomposition and heuristic\\ blockchain-enabled cooperative method to jointly \\optimize task offloading decisions \\and resource allocation.}
        & \makecell[l]{Proposed algorithm reduced delay by up to $40\%$\\ compared to non-cooperative method.\\ It is robust to increase in average task size\\ compared to other methods.}
        \\
        \hline        
        \cite{9113735} & \makecell{Blockchain-based \\MEC servers + Users} & \makecell{Secure resource allocation \\ + Max. system's efficiency} & 
        \makecell[l]{Decentralized blockchain-driven methods for\\ resource allocation in MEC, a.k.a, DAMB \\and BFDA.}
        & \makecell[l]{DAMB and BFDA mechanisms guarantee the \\truthfulness, individual rationality, and budget balance.\\Blockchain integration into resource allocation \\enhances the security and reliability of operations.} \\
        \hline
        \cite{10436712} & 
        \makecell{MEC servers \\+ BSs + Users}
        & \makecell{Max. utility function \\(task execution latency \\and energy conservation)}
         & \makecell[l]{Game theory based-hybrid task\\ offloading and scheduling algorithm.}& \makecell[l]{Proposed method outperformed benchmarks, in \\terms of task execution latency and energy conservation.}
        \\
        \hline
        \cite{10288204} & \makecell{MEC-equipped BSs \\+ Small Shore BSs \\+ Water surface users}
         & \makecell{Min. communication time \\from water surface \\ + Min. FL training time}
          & \makecell[l]{FL-based semantic communication\\ framework.}
          & \makecell[l]{Proposed FL optimization approach achieves \\fast convergence and high accuracy.\\ Communication bottleneck is solved through optimizing\\ resource allocation for underwater data transmissions.} \\
        \hline
        \cite{10453323} & \makecell{EH-equipped coastal BSs \\+ EH- and MEC-equipped \\Maritime stations + Users}
        & \makecell{Max. average \\network throughput} & \makecell[l]{Lyapunov-based JCORA for joint \\computation offloading and resource allocation. 
        } & \makecell[l]{Proposed method achieves a tradeoff between \\network throughput and queue stability.\\It outperforms other benchmarks.\\ 
        }
        \\
        \hline
    \end{tabular}%
    }
\end{table*}

Despite advances in ground infrastructure such as cellular networks, terrestrial networks still face significant challenges in offloading tasks. While advanced 5G technologies such as ultra-dense heterogeneous networks, massive MIMO, and millimeter wave communication have significantly improved data rates and reduced latency, task offloading remains a key issue. Mobile edge computing (MEC)\cite{8100873} addresses this problem by offloading tasks to edge nodes such as base stations, thereby reducing latency and improving performance. However, edge servers' limited capacity requires efficient computation and communication resource optimization. The following studies explore these challenges in more detail.  

Wang \textit{et al.} proposed in \cite{8493145} ``EdgeFlow'', a multilayer data flow processing system. They first qualified the system into two states: Blocking (data is buffered) and non-blocking (data is processed immediately). The framework aims to minimize system latency in non-blocking states and reduce buffer clearing time in blocking states. They used the Cauchy-Schwarz inequality to formulate optimization problems and designed algorithms for task assignment and resource allocation. Simulations showed EdgeFlow's robustness and efficiency in optimizing task assignment and resource allocation based on system state, outperforming other schemes. Teng \textit{et al.} presented in \cite{8732365} a Mixed-timescale Joint Computational Offloading and Wireless resource allocation (MJCW) algorithm for MEC systems to minimize energy consumption. The algorithm decouples the problem into short-term power and subcarrier allocation, and long-term task offloading and frequency scaling. This approach achieves significant energy savings and enhances green energy utilization. By managing energy resources efficiently and adjusting the energy price of MEC servers, the algorithm optimizes the use of green energy, leading to more sustainable MEC operations.
Also, authors of \cite{9956845} introduced an improved MEC and digital twin (DT)-based system called Wireless Computing Power Network (WCPN) for wireless networks like 6G. They proposed using computing resources from a pool of MECs and all devices with computing capabilities, termed "Vertical and Horizontal Scheduling." They aimed to minimize execution latency and energy consumption by implementing a task offloading algorithm based on multi-agent DRL and digital twin technology. The quick decision-making of the DRL model enabled the WCPN model to achieve high utility, while the number of iterations and participating end devices further enhanced its performance.
Moreover, Li \textit{et al.} proposed a Joint Resource Allocation and Computation Offloading (JRACO) scheme to optimize a millimeter-wave-based train-to-ground communication system \cite{9803864}. They considered the deployment of mobile relays (MRs) on trains operating in full-duplex mode. Its main objective was to minimize the average task execution latency for all users by optimizing resource allocation and computation offload strategies. By effectively addressing the combined challenge of allocating communication and computation resources, JRACO improves system performance, reducing latency and increasing the number of users served compared with baseline schemes. Similarly, Xu \textit{et al.} presented in \cite{10507238} a collaborative edge computing framework composed of multiple nodes. Their framework was blockchain-based, with computing node selection based on reputation-based consensus. They used the Tammer decomposition method and a heuristic approach to solve the joint optimization problem of task offloading and resource allocation. They utilized an improved NSGA-II algorithm to tackle the nonconvex optimization problem. Compared to greedy and non-cooperative algorithms, their proposed algorithm demonstrates a reduction in total delay of up to $40 \%$   when task size and transmission power are varied. 

By integrating blockchain, authors of \cite{9113735} prioritized the problem of computing resource allocation. They introduced two innovative double auction mechanisms for efficient resource allocation in blockchain-based MEC: the breakeven-based Double Auction Mechanism (DAMB) and the more efficient Breakeven-Free Double Auction Mechanism (BFDA). They proposed a Delegated Proof of Stake (DPoS) blockchain to guarantee a decentralized, secure, and fair consensus on resource allocation within MEC systems. The simulation results demonstrated that the DAMB and BFDA mechanisms significantly improve the efficiency of the MEC system, providing a more efficient and reliable computing environment for mobile users. In \cite{10436712}, the task offloading problem was formulated as a task offloading game to maximize both energy consumption and task execution utilities. The authors proposed a distributed hybrid task offloading algorithm for efficient coordination of various types of tasks. Using game theory, they modeled task offloading in an MEC-based framework and introduced an algorithm to find the Nash equilibrium. They also developed a priority-based task-scheduling algorithm. Their framework effectively used all available resources to achieve the lowest latency and highest energy utility.

In contrast to previous works, Picano \textit{et al.} extended the focus beyond ground-based computing to include water surface systems \cite{10288204}, as illustrated in Fig. \ref{fig:schema33}. They proposed a hybrid ground-water computing system by integrating edge infrastructure on the ground with shore-based stations. Their approach utilized semantic communication to minimize data retrieval time from water surface devices. The encoder was structured with a Conv2D stack and max-pooling layers, while the decoder employed upsampling layers. This framework has allowed them to reduce the convergence time of their FL models while maintaining high transmission accuracy. Fig.
Similar to \cite{10288204}, Wang \textit{et al.}  studied a MEC-powered water surface system \cite{10453323}. They developed a maritime MEC framework (composed of coastal BSs and MEC-enabled maritime information stations) with energy harvesting (EH) to enable sea lane monitoring. Their goal was to maximize average throughput under constraints of queue stability and energy budget. Their approach employed the Lyapunov method to solve the stochastic maximization problem and introduced the JCORA algorithm for joint optimization of task offloading over time and computing resource allocation. Simulations showed that JCORA achieved the highest average throughput when compared to benchmark schemes, highlighting its adaptability to dynamic sea environments. 
The aforementioned works are summarized in Table \ref{tab:ground}.

\Figure[t!](topskip=0pt, botskip=0pt, midskip=0pt)[width=0.95\columnwidth]{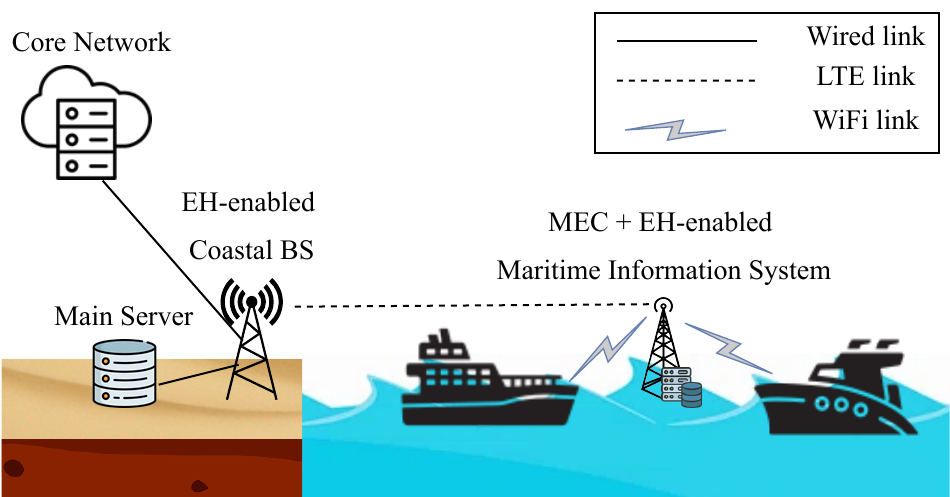}
{ \textbf{Network model \cite{10288204}.}\label{fig:schema33}}

\subsection{Vehicular Network-based Computing}
\begin{table*}[ht]
    \centering
    \Huge
    \caption{Related Work on Computing in Vehicular Networks}
    \label{tab:vehicular}
    \resizebox{\textwidth}{!}{%
    \begin{tabular}{|c|c|c|l|l|}
        \hline
        \textbf{Ref} & \textbf{System Model} & \textbf{Objective} & \textbf{Proposed Solutions} & \textbf{Findings} \\
        \hline
        \cite{8931659} & \makecell{SDN-based FiWi \\ + MEC-equipped RSUs \\+ Vehicle users} & \makecell{Min. processing delay} & \makecell[l]{ALBOA, GTNOA, PGTOA: Load balancing \\task offloading schemes.} & \makecell[l]{All schemes ensure QoS by keeping processing\\ delays within permissible limits.\\ ALBOA outperforms GTNOA and PGTOA by 20\%.} \\
        \hline
        \cite{9560096} & \makecell{MEC-equipped Vehicles \\ + Vehicle users} & \makecell{Min. task offloading latency} & \makecell[l]{Distributed fuzzy logic-based task offloading\\ scheme that leverages task decomposition into \\sub-tasks, processing parallelization, \\and  idle vehicular resources.} & \makecell[l]{Proposed approach outperforms existing schemes\\ in terms of latency, resource utilization, \\and successful offloading rate.} \\
        \hline
        \cite{9960947} & \makecell{MEC-equiped BSs \\ + Vehicle users} & \makecell{Min. overall latency \\and energy consumption} & \makecell[l]{Graph neural network for traffic prediction \\ and DDPG-based task offloading.} & \makecell[l]{In comparison with AAC and DPG, DDPG achieved\\ the lowest total latency and energy consumption.} \\
        \hline
        \cite{8955867} & \makecell{SDN-based \\ MEC-equipped RSUs \\ + Vehicle users} & \makecell{Max. system \\offloading utility} & \makecell[l]{JODTS: Hybrid task offloading\\ combining PGA and heuristic rules.} & \makecell[l]{JODTS achieved better offloading utility\\ and lower execution costs compared to benchmarks.
        } \\
        \hline
        \cite{9714172} & \makecell{MEC-equipped RSUs \\and Vehicles \\+ Vehicle users} & \makecell{Min. system latency} & \makecell[l]{GES algorithm for offloading decisions and \\JAOBSS for computation resource allocation.} & \makecell[l]{Proposed GES+JAOBSS solution is superior to \\benchmarks in terms of average system latency \\when varying number of vehicles and computation\\ capacity.} \\
        \hline
        \cite{10024868} & \makecell{SDN-based cloud computing\\ + MEC-equipped RSUs \\+ Vehicle users} & \makecell{Max. system \\social welfare} & \makecell[l]{BARGAIN-MATCH: Bargaining game-based\\ incentive approach for resource allocation\\ and many-to-one matching for inter-server\\ task offloading.} & \makecell[l]{BARGAIN-MATCH realized the highest social \\welfare for varied time, and vehicles' \\number and speed.} \\
        \hline
        \cite{10379502} & \makecell{MEC-equipped RSUs + \\ Vehicle users} & \makecell{Min. weighted energy \\ consumption} & \makecell[l]{Dynamic programming-based resource \\allocation and task offloading, and proximal \\gradient-based speed adjustment algorithm.} & \makecell[l]{The proposed algorithm has the lowest weighted \\ energy consumption in different scenarios e.g., task \\ size, intensity, number of vehicles, and latency limit.} \\
        \hline
        \cite{10061366} & \makecell{Cloud computing \\+ MEC-equipped RSUs \\and Vehicles + Vehicle users} & \makecell{Min. system-wide\\ user utility} & \makecell[l]{Asynchronous A3C for collaborative task\\ computing and on-demand management.} & \makecell[l]{Proposed A3C-based method achieved the lowest\\ latency compared to random and exclusive\\ local computing.} \\
        \hline
        \cite{9724147} & \makecell{Ground IRS-aided\\ MEC-equipped RSUs \\ + Vehicle users} & \makecell{Max. computation \\throughput} & \makecell[l]{Dynamic task scheduling algorithm based on\\ greedy search.} & \makecell[l]{Successfully computing rate and successfully \\finish rate of proposed approach outperform \\those of random and  FIFO benchmarks.} \\
        \hline
        \cite{10007043} & \makecell{Cloud computing\\+ MEC-equipped RSUs + \\ Vehicle users} & \makecell{Min. average \\task latency} & \makecell[l]{DDPG-based method for \\ task assignement and distributed reactive caching.} & \makecell[l]{Proposed algorithm achieved the best latency,\\ and a trade-off between task offloading latency\\ and energy consumption.} \\
        \hline
    \end{tabular}%
    }
\end{table*}

The uprising paradigm of vehicular edge computing (VEC) has been a popular topic in research, especially with the incoming 6G-based ITS and vehicle-to-everything (V2X) communications \cite{10379502}. A VEC network is composed of edge computing servers, a.k.a., road-side units (RSUs), placed close to vehicles and computing resources mounted on the vehicles themselves, where vehicle-to-infrastructure (V2I, e.g., between vehicles and RSUs) or vehicle-to-vehicle (V2V) communications can be established. Nevertheless, vehicular applications such as autonomous driving, platooning, and infotainment may require high computing power to process large amounts of collected data. To overcome this challenge, several research efforts have been carried out. 

For instance, 
Zhang \textit{et al.} introduced in \cite{8931659} a software-defined network (SDN)-based framework for load balancing and task offloading for FiWi (Fiber-Wireless) enhanced VEC, to minimize the processing delay. They introduced two game theory-based offloading schemes: the Nearest Offloading Algorithm (GTNOA) and the Predictive Offloading Algorithm (PGTOA). These schemes enhance the decision-making process for task offloading among vehicles and edge servers. Also, they presented an approximate computing offloading scheme called ALBOA to achieve load balancing among the computation resources of the MEC. The results demonstrated the superior performance of the load balancing scheme in improving the user experience. Interestingly, ALBOA outperformed GTNOA and PGTOA as the total processing delay achieved is approximately 20\% lower compared to GTNOA and PGTOA.
In \cite{9560096}, Bute \textit{et al.} minimized the latency using a distributed task offloading scheme that leveraged idle vehicle computing resources for parallel task processing. Their fuzzy logic-based offloading scheme selects the optimal offloading targets based on link reliability, distance, and available computing resources. 
Xu \textit{et al.} designed in \cite{9960947} a vehicular computation offloading framework where they aimed to minimize the overall latency and energy consumption. They proposed a graph-weighted convolution network to efficiently predict traffic flow and allocate resources. 
The offloading decision is then determined using a deep deterministic policy gradient-based (DDPG) method. Experiments demonstrated the superiority of DDPG over Actor-Critic (AC) and Deterministic Policy Gradient (DPG)-based schemes.
In \cite{8955867}, the authors investigated an SDN-based task offloading framework for VEC, to maximize an offloading utility defined by task delay and computing resource consumption. To solve the formulated task offloading problem, they proposed a joint offloading decision and task scheduling algorithm (JODTS) inspired by the partheno genetic algorithm (PGA) and heuristic rules.


On the other hand, 
Feng \textit{et al.} proposed in \cite{9714172} a reverse offloading framework, called a cooperative vehicle-infrastructure system (CVIS), that leverages the vehicles' computing resources to support external VEC servers. Their objective is to minimize system latency through the optimization of task assignment, and resource allocation. Two offloading algorithms have been proposed, namely greedy-based efficient searching (GES) for task assignement and joint alternative optimization-based bi-section searching (JAOBSS) for resource allocation. 
Moreover, the authors in \cite{10024868} proposed a hierarchical resource allocation and task offloading mechanism for the SDN-based VEC network. They formulated a joint computation resource allocation and task offloading problem to maximize the system's social welfare. To deal with the NP-hardness of the problem, they used a bargaining game approach for resource allocation, meanwhile, task offloading was handled by a many-to-one matching approach for horizontal (edge-edge) and vertical (edge-cloud) migrations. Their solution, called BARGAIN-MATCH, maximized the system's social welfare, outperforming benchmarks. In \cite{10379502}, Li \textit{et al.} formulated the problem of computation offloading in speed-adjustable VEC to minimize the weighted energy consumption of tasks. They studied the correlation between task offloading and the moving speed of a vehicle, on top of proposing a dynamic programming-based task offloading and resource allocation strategy. They proposed a speed adjustment algorithm inspired by the proximal gradient to maintain an optimal vehicle location with low communication delays. 
Liu \textit{et al.} leveraged both V2V and V2I in their VEC resource allocation framework \cite{10061366}. Their formulated problem consists of minimizing the system-wide user utility through joint optimization of task assignment and resource scheduling. Their solution is based on the asynchronous advantage actor-critic (A3C) algorithm.
In contrast, the authors of \cite{9724147} leveraged intelligent reflecting surface (IRS)-assisted communications in VEC to maximize computation throughput. They proposed a greedy-based dynamic task-scheduling strategy that is proven to outperform random and FIFO benchmarks.
Finally, Xue \textit{et al.} integrated caching into computation offloading in an energy-aware VEC framework, aiming to reduce average task latency, as illustrated in Fig. \ref{fig:schemaVEC} \cite{10007043}. The formulated problem is solved using a DDPG-based method for task offloading and a distributed reactive caching placement strategy. The proposed solution minimized latency while achieving a trade-off between energy consumption and delay.  
These works are summarized in Table \ref{tab:vehicular} above.

\Figure[t!](topskip=0pt, botskip=0pt, midskip=0pt)[width=0.9\columnwidth]{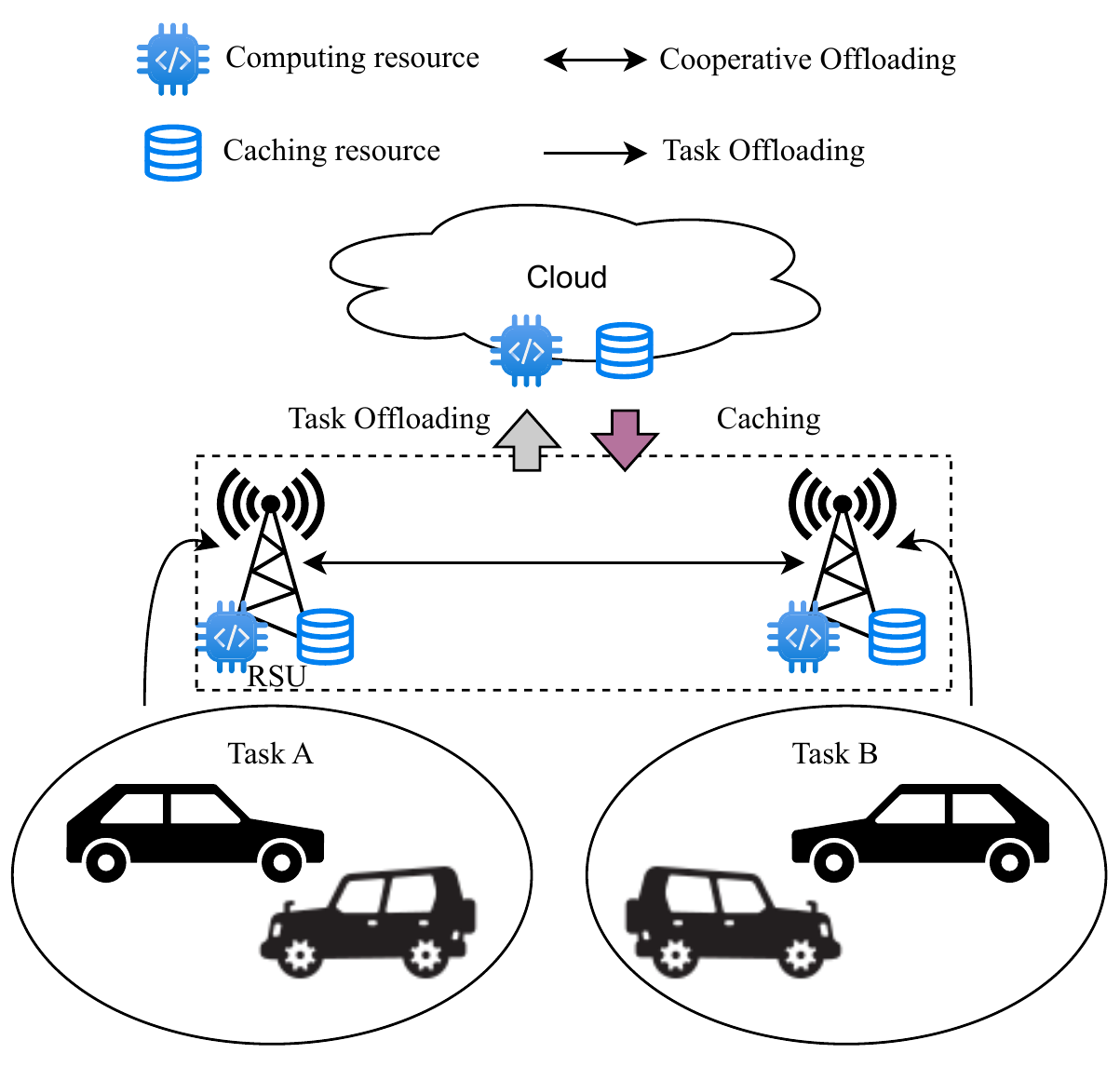}
{ \textbf{VEC framework in \cite{10007043}.}\label{fig:schemaVEC}}

\subsection{IoT Network-based Computing}

Given the massive number of IoT devices, a huge leap in data traffic at the edge of the networks has been experienced, surpassing the capacity of traditional networks, notably in providing sufficient bandwidth and satisfying latency. In response to this issue, IoT-based edge computing has emerged as a transformative solution. 
For instance, Liu \textit{et al.} investigated in \cite{9462411} task offloading in a MEC-supported IoT system with fiber backhaul, for which they proposed the optimized task allocation approach (OTAA), based on a biogeography-based optimization (BBO) algorithm, to minimize the unavailability of task allocations and resource wastage, while maintaining communication QoS in terms of delay.
In a similar IoT-edge-cloud system, authors of \cite{8736731} leveraged a workload allocation mechanism to minimize energy consumption, given delay limitations. 
Based on the Lyapunov drift-plus-penalty theory, the proposed delay-based workload allocation (DBWA) algorithm was used to define the optimal workload allocation decision between the local edge server, neighboring servers, and cloud servers. 
Alternatively, Deng \textit{et al.} leveraged the Lyapunov optimization method to develop the dynamic parallel computing offloading and energy management (DPCOEM) algorithm in \cite{8854900}. DPCOEM addresses the minimization problem of response time and packet loss of energy harvesting (EH)-based MEC IoT networks.  
In \cite{8648197}, the authors aimed at minimizing the long-term average weighted sum of delay and power consumption under stochastic traffic arrival in a Narrowband-IoT (NB-IoT) MEC system. The authors jointly formulated the problem of computation offloading and radio resource allocation using continuous-time MDP (CTMDP). Linear value function approximation and temporal-difference learning method were then employed to derive an RL-based approximate dynamic programming (ADP) algorithm for computation offloading and user scheduling. 
Authors of \cite{9060882} leveraged also RL for resource allocation in MEC-supported IoT networks, to reduce the average completion time of tasks and average requested resources. 
MDP is used to formulate the problem and it is solved by an improved DQN algorithm. A new filter layer is added at the end of DQN to remove the values of invalid actions. While the policy initially showed higher completion times, it consumed fewer resources over time as the model is being trained to achieve a trade-off between completion time and number of requested resources. 
Differently from the previous approaches, a multi-IRS-assisted MEC-system for IoT was studied in \cite{9881553}. The objective is to maximize the sum computation rate (SCR) for single and multiple EH IoT users, in scenarios of partial (part of the task only) and full (whole task) task offloading.
For partial offloading, the alternating optimization (AO) method was used to optimize jointly the IRS phase shifting and partial task assignment, while a simpler greedy algorithm was proposed in the full offloading scenario. 

Joint optimizations of task offloading and computation resource allocation have been conducted in the works below. First, 
Fan \& Ansari proposed in \cite{8336866} an application-aware workload allocation scheme (AREA) to minimize the average response time of SDN and cloudlet-based IoT systems. They proposed sequential solving where the task assignment strategy is based on a heuristic algorithm, and then the resource allocation convex problem per cloudlet is solved with typical convex optimization tools.
Similarly, Shah-Mansouri and Wong investigated in \cite{8360511} joint task offloading and computing resource allocation in a cloud-fog-supported IoT system. To maximize the quality of experience (QoE) of IoT users, translated into a minimization function of computation energy and delay, they developed the task offloading strategy based on graph theory, while Nash Equilibrium was used to achieve near-optimal resource allocation.
In \cite{9134385}, a dual-layer MEC-supported IoT network has been studied, where the objective is to minimize the system's latency and energy consumption. They proposed a joint task assignment and resource allocation solution based on a heuristic algorithm for assignment and an improved discrete particle swarm optimization (PSO) for resource allocation.  
Xia \textit{et al.} explored in \cite{9424444} a distributed MEC-supported IoT system that incorporates multiple edge-cloud servers and EH-enabled IoT users. The objective is to maximize the utility of the system over time with respect to energy consumption and computing latency. They proposed an online distributed optimization algorithm based on buyer-seller game theory and perturbed Lyapunov optimization, thus achieving optimal task offloading and computing resource allocation. 
In \cite{9740226}, the authors proposed a multi-step computation offloading in a multi-BS supported IoT network. They defined one-step offloading to macro BSs only, while two-step offloading is to both small and macro BSs. The goal is to minimize network energy consumption through BS-IoT user association, power control, task assignment, and frequency band partitioning, under latency constraints. They proposed an improved hierarchical adaptive search (IHAS) algorithm, combining a genetic algorithm (GA) and an improved adaptive PSO (IAPSO). 
Unlike previous works, Kim \textit{et al.} considered in \cite{9276401} a MEC-supported IoT network (see Fig. \ref{fig:schemaIoT}) where IoT devices can process tasks locally and assist edge servers, thus enabling a better usage of idle IoT resources.
\Figure[ht!](topskip=0pt, botskip=0pt, midskip=0pt)[width=0.85\columnwidth]{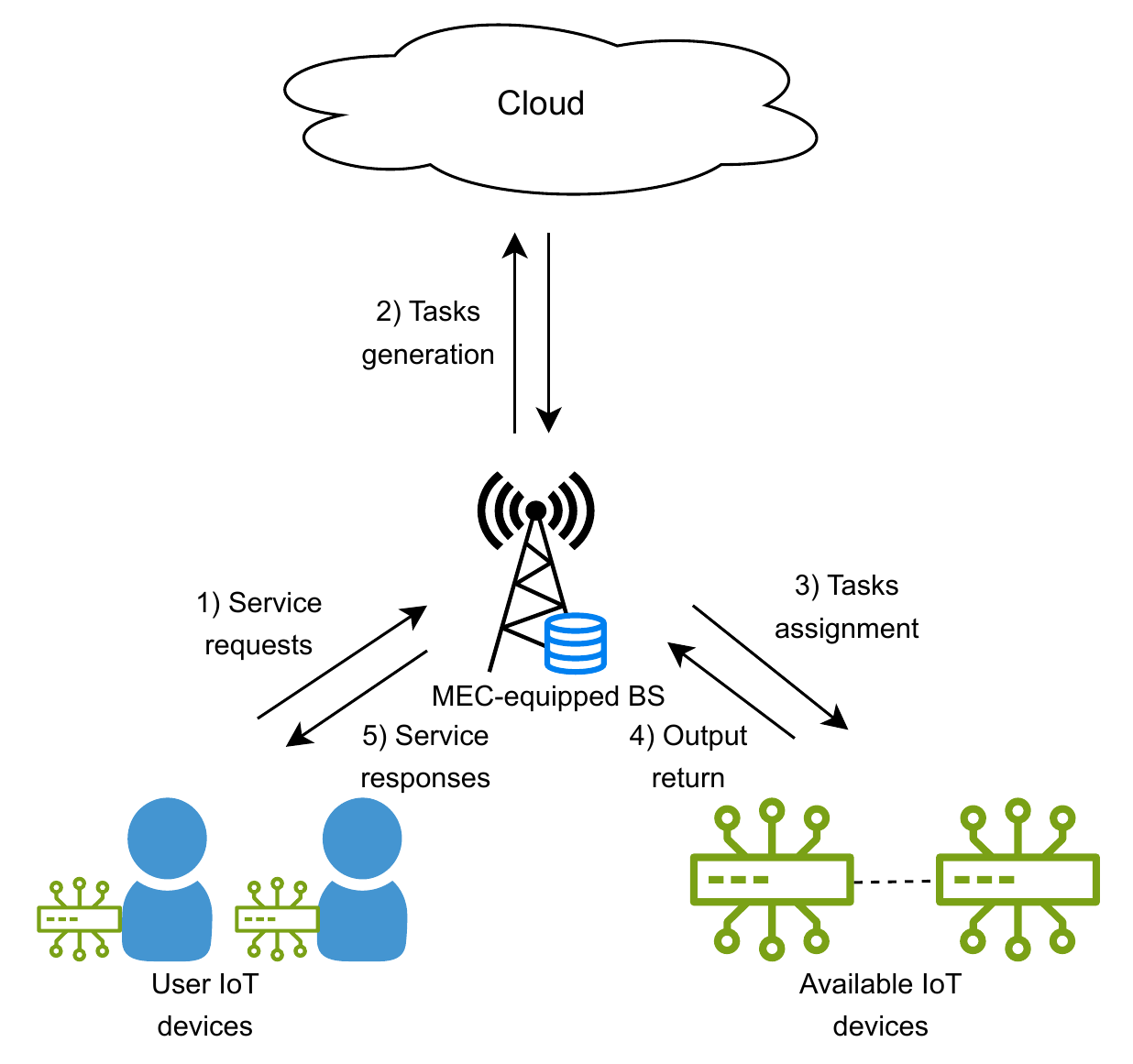}
{ \textbf{System model in \cite{9276401}.}\label{fig:schemaIoT}}
To solve the task assignment problem, they developed a completion time-based task assignment (CTA) algorithm, and for scheduling, a hierarchical weight allocation (HWA) algorithm. Simulations showed that HWA ensured that all tasks met their deadlines, while CTA improved the likelihood of critical tasks meeting their own. All the aforementioned works are summarized in Table \ref{tab:iot}.

\begin{table*}[ht]
    \large
    \centering
    \caption{Related Work on Computing in IoT Networks}
    \label{tab:iot}
    \resizebox{\textwidth}{!}{%
    \begin{tabular}{|c|c|c|l|l|}
        \hline
        \textbf{Ref} & \textbf{System Model} & \textbf{Objective} & \textbf{Proposed Solutions} & \textbf{Findings} \\
        \hline
        \cite{9462411} & \makecell{Cloud computing\\+ MEC-equipped BSs\\ + IoT users} & \makecell{Min. unavailability \\of task allocations \\and resource wastage} & \makecell[l]{OTAA: Optimized task allocation\\ based on BBO.}
        & \makecell[l]{OTAA reduced task allocation unavailability and \\resource wastage compared to random, PSO, and \\first-fit baselines.\\Number of BSs and IoT significantly impacts\\ performances.}
        \\
        \hline
        \cite{8736731} & \makecell{Cloud computing\\+ MEC servers\\ + IoT users} & \makecell{Min. energy consumption  
        } & \makecell[l]{DBWA: Lyapunov drift-plus-penalty theory-based\\ task assignment to cloud and edge servers.} & \makecell[l]{DBWA achieves the lowest energy consumption\\ while maintaining delay below threshold compared\\ to full MEC and full cloud offloading benchmarks.} \\
        \hline
        \cite{8854900} & \makecell{MEC servers\\ + MEC-equipped \\EH IoT + EH IoT users} & \makecell{Min. response time \\and packet loss} & \makecell[l]{DPCOEM: Dynamic parallel task offloading\\ and energy management based on Lyapunov\\ optimization.} & \makecell[l]{
        DPCOEM minimizes average task cost, \\completion time, and ratio of dropped tasks, \\in different conditions of task arrival\\ probability, average EH power, and deadlines.} \\
        \hline
        \cite{8648197} & \makecell{MEC-equipped \\NB-IoT BS \\+ IoT users} & \makecell{Min. avg. \\weighted sum delay\\ and power consumption} & \makecell[l]{ADP: RL algorithm for joint computation\\ offloading and user scheduling.} & \makecell[l]{ADP outperforms baseline algorithms in terms\\ of the weighted sum of average delay and power \\consumption for any IoT network size.} \\
        \hline
        \cite{9060882} & \makecell{Cloud computing \\ + MEC-equipped BSs \\ + IoT users} & \makecell{Min. avg. completion time \\  and avg. requested resources} & \makecell[l]{Enhanced DQN-based algorithm for \\resource allocation.} & \makecell[l]{Proposed algorithm achieves a lower loss than \\conventional DQN algorithm. A trade-off between \\task completion and number of requested \\computing units is identified. 
         } \\
         \hline
         \cite{9881553} & \makecell{MEC-equipped BS\\ + Multiple IRSs \\+ EH IoT users} & Max. SCR & \makecell[l]{AO method including discrete phase\\ shifting using successive convex appropriation\\ (SCA) and one-dimensional search for task \\scheduling in time. For partial offloading \\ a greedy algorithm for binary offloading} & \makecell[l]{
         AO-based partial offloading achieves \\better SCR than greedy-based full\\ offloading. Multi-IRS system improves\\ SCR compared to without-IRS system.} \\
        \hline
        \cite{8336866} & \makecell{SDN-based cloud \\computing + Cloudlets \\ + IoT users} & \makecell{Min. avg. response time} & \makecell[l]{AREA: Heuristic scheme for task assignment \\ Convex optimization for resource allocation.} & \makecell[l]{AREA realizes low response time compared\\ to baselines for different parameter values,\\ e.g., capacity of cloudlets, number of users,\\ and workload.} \\
        \hline
        
        \cite{8360511} & \makecell{Cloud computing \\+ Fog servers\\ + IoT users} & Max. QoE & \makecell[l]{Nash equilibrium game for near-optimal \\resource allocation.} & \makecell[l]{
        Proposed algorithm improves QoE of more\\ IoT users compared to baselines. Addition of \\fog nodes reduces computation time by 70\%.} \\
        \hline

        \cite{9134385} & \makecell{Cloud computing\\ + two-layer MEC \\servers + IoT users} & \makecell{ Min. avg. task latency \\and energy consumption} & \makecell[l]{Heuristic dynamic task assignment and improved\\ discrete PSO algorithm for resource allocation.} & \makecell[l]{Proposed approach reduces latency, bandwidth \\usage, and energy consumption.} \\
        \hline
        \cite{9424444} & \makecell{MEC servers\\ + EH IoT users} & \makecell{Max. network\\ utility} & \makecell[l]{Online distributed optimization algorithm\\ based on buyer-seller game and perturbed\\ Lyapunov optimization .} & \makecell[l]{Proposed scheme is superior to baselines\\ in terms of network utility.} \\
        \hline
        \cite{9740226} & \makecell{Macro BSs \\+ Small BSs\\ + IoT users} & \makecell{Min.\\ energy consumption} & \makecell[l]{IHAS: Hybrid algorithm based on GA and \\IAPSO for joint BS-IoT user association,\\ power control, task assignment,\\ and frequency band partitioning.} & \makecell[l]{IHAS outperforms baselines in terms\\ of energy consumption.} \\
        \hline
        \cite{9276401} & \makecell{MEC servers\\ + MEC-equipped IoT \\+ IoT users} & \makecell{Min. completion time \\ max. satisfaction rate.} & \makecell[l]{CTA for task scheduling and HWA \\for computation resource allocation. } & \makecell[l]{HWA guarantees that the execution of\\ tasks respects deadlines. CTA ensures that\\ a high probability of critical tasks \\meet their deadlines.} \\
        \hline
    \end{tabular}%
    }
\end{table*}

\section{Computing in Non-Terrestrial Networks}\label{sec:ntn}

NTNs effectively complement terrestrial networks by providing a wide area coverage, ensuring connectivity in under-served and hard-to-reach regions, such as rural areas, maritime environments, and disaster-stricken areas. In addition, NTNs can provide edge services, bringing computing resources and data processing closer to end users. According to the 3rd Generation Partnership Project (3GPP), an NTN can be composed of spaceborne systems, e.g., geostationary, medium, and low Earth orbit satellites (GEO, MEO, and LEO) or airborne systems, e.g., UAVs and HAPS.
In what follows, we provide a comprehensive understanding of each component of NTN and its integration with edge computing.

\subsection{Satellite Network-based Computing}

Satellite networks have been around for communications since the 1950s, providing wide connectivity to Earth users. 
In recent years, attention has shifted towards the development of LEO systems mainly for their low link propagation latency, compact size, and longer life span \cite{9372909}. Moreover, LEO satellites can host MEC servers, thus enabling the processing of massive data. 
For instance, in \cite{10107794}, a satellite peer offloading scheme was introduced to jointly maximize the LEO utilization of computing resources and minimize energy consumption while considering resource and backlog constraints. The authors formulated the offloading problem in a collaborative multi-hop system, called multi-hop satellite peer offloading (MHSPO), and solved it using the Lyapunov method 
and an online queue-based distributed offloading scheme. Their approach achieved the lowest latency and second lowest energy consumption compared to single-hop benchmarks. 
The authors of \cite{10439163} focused on minimizing service delay in a satellite edge computing (SEC) system by optimizing the selection of offloading paths between satellites, taking into account energy and storage resources limitation. The optimization problem is formulated using MDP, then solved using a new dueling double deep Q-network (D3QN)-based offloading scheme. 
A distinction in the type of tasks is proposed with the greedy-based orbital edge computing task allocation (OEC-TA) scheme in \cite{9894082}. The latter leveraged task classification into delay-sensitive (DST) and delay-tolerant tasks (DTT), and used queueing theory to calculate and optimize each task's response time and energy consumption. 
Also, Tang \textit{et al.} developed in \cite{10436113} a two-timescale hierarchical framework to address task offloading and service deployment, aiming to minimize energy consumption, load imbalance, and packet loss. On small timescales, the task offloading overtime problem is first formulated using constrained MDP (CMDP) and then solved by a soft actor-critic (SAC)-based method. On large timescales, integer programming is used to formulate the service deployment problem, and a low-complexity heuristic atomic orbital search (AOS) approach was developed to solve it. 
Similar to \cite{10436113}, the authors of \cite{10266782} proposed an approach that optimizes the simultaneous placement of edge servers and service provisioning in a satellite-terrestrial integrated edge computing network (STECN), targeting the minimization of the average response latency and energy consumption. Their method relied on K-medoids clustering for servers placement and NSGA-II heuristic for service deployment. 
In \cite{9978929}, the authors proposed a MEC-based satellite-terrestrial framework with one LEO and multiple ground BSs that offload users' tasks, aiming to minimize the system’s energy dissipation while maintaining a low computing latency. To solve the formulated NP-hard problem, they proposed an alternating optimization (AO)-based method that alternatively optimizes the communication and computing resource allocation.   
Similarly, 
Zhang \textit{et al.} proposed in \cite{10032271} a DDPG-based computation offloading approach to minimize energy consumption under resources and delay constraints, in a three-tier MEC-based network (cloud, satellites, and IoT users). 
In \cite{10227270}, the authors' objective was to minimize the long-term power cost of a MEC-equipped LEO-based network, by jointly optimizing task offloading and resource allocation. Through decomposition into sub-problems and mixing DRL with conventional optimization, a.k.a., DRLCO, they solved the mixed integer non-linear programming (MINLP) problem. 
Moreover, Hassan \textit{et al.} considered in \cite{10198334} a MEC-enabled satellite network for ITS services. They jointly optimized offloading task assignment, and computing and bandwidth resource allocation, to minimize delay and rental price of satellite servers. They proposed a two-stage algorithm, based on cooperative multi-agent proximal policy optimization (Co-MAPPO) with attention and convex theory, to solve it. 
In \cite{8675467}, the authors proposed a software-defined satellite-terrestrial network for the joint dynamic orchestration of caching, networking, and computing resources. 
The objective is to increase the system's utility per resource by optimizing caching, MEC, and networking resource allocation. 
Based on an MDP representation, a deep Q-network (DQN) scheme has been developed to jointly optimize the networking, caching, and edge computing resources usage. 
Hao \textit{et al.} studied the joint optimization of computation offloading, radio resource allocation, and caching placement in a MEC-enabled LEO network, to reduce the tasks' total delay \cite{10018385}. The proposed Lagrange dual decomposition (LDD)-based algorithm solves the formulated problem, by deriving a closed-form solution. The latter's complexity is reduced using a heuristic algorithm. 
Finally, Huang \textit{et al.} developed in \cite{10629187} a STECN dual-timescale optimization framework to reduce energy consumption and task processing delay. At a high time frequency, task offloading is optimized using 
a DRL-based solution, while at a lower time frequency, a heuristic-based artificial electric field (AEF) method is adopted to optimize resource allocation.
A self-attention mechanism is also introduced to allow dynamic information exchange between satellites regarding load-balancing and resource allocation. 
Table \ref{tab:stn} provides a summary of the above works.

\begin{table*}[ht]
    \centering
    \huge
    \caption{Related Work on Computing in Satellite Networks}
    \label{tab:stn}
    \resizebox{\textwidth}{!}{%
    \begin{tabular}{|c|c|c|l|l|}
        \hline
        \textbf{Ref} & \textbf{System Model} & \textbf{Objective} & \textbf{Proposed Solutions} & \textbf{Findings} \\
        \hline
        \cite{10107794} & \makecell{MEC-equipped Sat. \\ + Users} & \makecell{Max. Sat. computing \\resource usage\\ Min. energy consumption} & \makecell[l]{Online distributed decision-making using \\Lyanupov technique and queueing to solve \\MHSPO problem.} & \makecell[l]{Proposed solution achieves the best user satisfaction\\ user with minimum delay. \\ Energy consumption is slightly higher than in the \\single-hop benchmark system.} \\
        \hline
        \cite{10439163} & \makecell{MEC-equipped \\LEO Sat. + Users} &\makecell{ Min. service delay} & \makecell[l]{D3QN-based task offloading with\\ service migration.} & \makecell[l]{D3QN has the lowest average service delay and\\ convergence time compared to benchmarks, for\\ any satellite storage and energy capacity parameters.} \\
        \hline
        \cite{9894082} & \makecell{MEC-equipped \\LEO Sat. + Users} & \makecell{Min. computing cost, delay, \\ and energy consumption} & \makecell[l]{Greedy-based OEC-TA algorithm, with distinction \\and prioritization between DST and DTT.} & \makecell[l]{OEC-TA reduces delay and energy cost of task\\ execution in both DST and DTT by 10\% and 16.5\%\\ compared to ``double edge computing'' benchmark.} \\
        \hline
        \cite{10436113} & \makecell{Cloud computing \\ + MEC-equipped \\LEO Sat. + Users} & \makecell{ Min. energy consumption, \\load imbalance,\\ and packet loss} & \makecell[l]{SAC-based method for task offloading decisions.\\ AOS heuristic for service deployment.} & \makecell[l]{Proposed scheme reduces average \\ task scheduling cost by 39.4\% compared \\to the stationary deployment scheme.} \\
        \hline
        \cite{10266782} & \makecell{MEC-equipped \\LEO Sat. + Users} & \makecell{Min. response latency \\and energy consumption} & \makecell[l]{K-medoids clustering for servers placement.\\ NSGA-II heuristic for service deployment.} & \makecell[l]{Convergence of proposed NSGA-II algorithm \\is achieved in 50 iterations. Proposed clustering \\reduces total cost by up to 63.85\% and latency \\by 53.79\% compared to benchmarks. 
        } \\
        \hline
        \cite{9978929} & \makecell{MEC-equipped \\LEO Sat. \\+ BSs + Users} & \makecell{Min. system's \\energy dissipation} & \makecell[l]{AO approach based on SDP and Lagrangian \\multipliers for task offloading and resource \\allocation optimization, respectively.} & \makecell[l]{Proposed approach has the best energy consumption\\ and system delay when changing user computational \\capacity, number of users, number of BSs.} \\
        \hline
        \cite{10032271} & \makecell{Cloud computing \\ + MEC-equipped \\LEO Sat. + IoT users} & \makecell{Min. energy consumption} & \makecell[l]{DDPG-based algorithm for task offloading.} & \makecell[l]{Proposed method achieves the lowest latency and \\energy consumption compared to benchmarks, and \\for different task arrival rates and computing capacity.} \\
        \hline
        \cite{10227270} & \makecell{MEC-equipped \\LEO Sat. + Users} & \makecell{Min. long-term \\ power cost} & \makecell[l]{DRLCO, a four-stage method that jointly\\ optimizes task offloading and communication \\and computing resource allocation.} & \makecell[l]{DRLCO has the lowest average total power \\ cost compared to DQN and heursitic baselines.} \\
        \hline
        \cite{10198334} & \makecell{MEC-equipped GEO, \\ LEO and cube Sat. \\ + ITS users} & \makecell{Min. offloading delay \\and rental price \\of Sat. servers} & \makecell[l]{Two-stage algorithm based on Co-MAPPO \\algorithm and attention mechanism for \\intelligent task offloading decisions.} & \makecell[l]{
        Proposed method balances between offloading\\ delay and price to achieve the best objective value,\\ compared to benchmarks.} \\
        \hline
        \cite{8675467} & \makecell{SDN-based LEO Sats.\\ + MEC servers\\ + Caches + Users} & \makecell{Max. expected utility \\per resource} & \makecell[l]{DQN-based method for joint allocation of \\networking, caching, and computing resources.} & \makecell[l]{DQN realizes the highest utility per resource \\compared to benchmarks for different satellite \\elevations, content sizes, and resource prices.} \\
        \hline
        \cite{10018385} & \makecell{MEC/Cache-equipped\\ LEO Sats. + IoT users} & \makecell{Min. tasks \\total delay} & \makecell[l]{LDD-based method and a lower complexity heuristic \\for joint optimization of computation offloading,\\ radio resource allocation, and cache placement.} & \makecell[l]{Proposed algorithms outperform full local \\and full-MEC baselines, in terms of delay.} \\
        \hline
        \cite{10629187} & \makecell{MEC-equipped \\LEO Sat. + Users} & \makecell{Min.  energy consumption\\ 
        and task processing delay} & \makecell[l]{Dual-timescale optimization: DRL-based method \\for task offloading (frequent), and AEF heuristic for \\resource allocation (less frequent).} & \makecell[l]{Proposed scheme achieves lowest processing\\ cost (up to 15\% less) and highest energy efficiency\\ (up to 52\% more) compared to baselines.}\\
        \hline
    \end{tabular}%
    }
\end{table*}

\subsection{HAPS Network-based Computing}
The development of HAPS has been long coveted as a key enabling technology of beyond 5G (B5G) and 6G. Its high flexibility and low deployment cost, compared to satellites, provides it with an advantage to achieve short communication latency, wide coverage, and significant computing service support. HAPS can operate as a medium to coordinate fast and reliable communications between satellites, reducing handovers, managing UAV swarms, supporting IoT and Internet-of-vehicle (IoV) systems, and provisioning Internet access to ground users \cite{10045716,10186454}. In the following, we present recent works that combined HAPS and MEC to support user tasks.


HAPS has been proposed to support vehicular services. For instance, Shinde \textit{et al.} studied in  \cite{9650754} the task offloading problem in a HAPS-assisted VEC system. The HAPS, in their use case, is used as a federated learning (FL) server to evaluate computation offloading parameters, while CAVs offload their tasks to RSUs. An evolutionary GA scheme has been then proposed to solve the minimization problem of latency and energy consumption. 
Ren \textit{et al.} presented in \cite{10103832} a handoff-aware computing scheme that exploits the available resources in vehicles, HAPS, and RSUs, to minimize the task execution delay. 
They proposed an optimization policy for task-splitting, vehicle transmit power, bandwidth, and computing resources allocation, using variable replacement and successive convex approximation (SCA). Their method involves HAPS to address computing interruptions caused by handoffs and minimize latency.  
In \cite{10024371}, the authors proposed to leverage millimeter wave operating  HAPS-enabled VEC in rural areas. The goal was to maximize successful task offloading by optimizing the offloading factor of the task between the CAV and HAPS, under latency and computation capacity constraints. They optimally solved the formulated problem using the Brent solvers.
Authors of \cite{9931920} proposed a heuristic-based adaptive task offloading strategy to minimize the processing time in a HAPS-based MEC system. 
By accounting for task properties, dependencies, and decomposability, their approach dynamically selects offloading hosts and thus reduces the task processing time. 

In \cite{9714482}, the authors introduced a hierarchical aerial computing network with HAPS, UAVs, and terrestrial IoT users. They studied the problem of maximizing the total amount of IoT data processed by aerial MEC platforms while satisfying the delay QoS and resource limitations at HAPS and UAVs.
Given its complexity, this problem is solved in two stages, with the first consisting of offloading data from IoT devices to UAVs using a matching game-based scheme, while the second uses a heuristic algorithm for task offloading between UAVs and HAPS. 
In addition, the authors in \cite{9430617} targeted minimizing energy and time consumption for computation and task transmission in HAPS-based networks. To achieve this, they jointly optimized user association, service sequence, and task allocation using a novel support vector machine (SVM)-based FL algorithm. 
Abderrahim \textit{et al.}  
\begin{table*}[t]
    \centering
    \huge
    \caption{Related Work on Computing in HAPS Systems}
    \label{tab:haps}
    \resizebox{\textwidth}{!}{%
    \begin{tabular}{|c|c|c|l|l|}
        \hline
        \textbf{Ref} & \textbf{System Model} & \textbf{Objective} & \textbf{Proposed Solutions} & \textbf{Findings} \\
        \hline
        \cite{9650754} & \makecell{HAPS (FL server) \\ + RSUs + Vehicle users} & \makecell{Min. overall latency\\ and energy consumption} & \makecell[l]{FL for computing offloading parameters \\evaluation and evolutionary GA for \\task offloading.} & \makecell[l]{Proposed method achieves the lowest latency \\and energy consumption, compared \\to benchmarks.} \\
        \hline
        \cite{10103832} & \makecell{MEC-equipped HAPS \\+ RSUs + Vehicle users} & \makecell{Min. task delay} & \makecell[l]{
        SCA-based method for joint optimization of task \\splitting, CAV transmit power, bandwidth, \\and computing resources.
        } & \makecell[l]{Proposed approach reduced failed workloads\\ and latency, and maintained system stability.} \\
        \hline
        \cite{10024371} & \makecell{MEC-equipped HAPS \\+ Vehicle users} & \makecell{Max. successful \\task offloading} & \makecell[l]{Brent solvers for offloading factor optimization.} & \makecell[l]{Proposed optimized offloading achieves \\the lowest average latency and highest \\real-time service probability,  outperforming \\full local computing.} \\
        \hline
        \cite{9931920} & \makecell{MEC-equipped HAPS \\and BSs + Users} & \makecell{Min. total computing\\ time of tasks} & \makecell[l]{Task decomposition and greedy offloading based \\on task inter-dependencies and characteristics.} & \makecell[l]{Proposed algorithm achieves the lowest processing\\ time compared to local processing at user or BS.} \\
        \hline
        \cite{9714482} & \makecell{MEC-equipped HAPS \\ and UAVs + IoT users} & \makecell{Max. task \\satisfaction rate} & \makecell[l]{Matching game theory-based algorithm for IoT \\task offloading to UAVs, coupled with a 
        heuristic \\method for offloading from UAVs to HAPS.} & \makecell[l]{Proposed solution improves task satisfaction rate,\\ beyond benchmarks.} \\
        \hline
        \cite{9430617} & \makecell{MEC-equipped \\multiple HAPS + Users} & \makecell {Min. weighted sum\\ of energy and time\\ consumption} & \makecell[l]{SVM-based FL algorithm for joint optimization\\ of user association, service sequence, \\and task allocation} 
        & \makecell[l]{SVM-based FL reduced the weighted sum of \\energy and time consumption up to 15.4\% \\compared to centralized benchmarks.
        } \\
        \hline
        \cite{10506092} & \makecell{MEC-equipped\\ HAPS and UAVs \\+ Users} & \makecell{Min. long-term \\processing cost} & \makecell[l]{
        Multi-agent DRL-based task offloading algorithm \\enhanced with ConvLSTM for UAV task load\\ estimation, and PER for accelerated training.} & \makecell[l]{Proposed method has an average cost\\ lower by up to 23.2\% compared to\\ the metaheuristic benchmarks. 
        } \\
        \hline
        \cite{10373079} & \makecell{MEC-equipped\\ HAPS and UAVs \\+ IoT users} & \makecell{Min. energy \\consumption} & \makecell[l]{DOTORA: Alternating optimization method that \\uses convex optimization for local computation \\resource allocation, DGMS for task offloading,\\ and TPA for power allocation. 
        }
        & \makecell[l]{DOTORA reduced energy consumption \\by up to 19\% compared to baselines.} \\
        \hline
        \cite{9515574} & \makecell{MEC-equipped \\LEO Sat. and multiple \\HAPS + Users} & \makecell{Min. weighted sum \\energy consumption} & \makecell[l]{QTFP method for user association. WMMSE for \\MU-MIMO precoding. Interior point method \\for task assignment. Closed-form expression \\for computation resource allocation.
        }
        & \makecell[l]{Proposed solution has the lowest energy\\ consumption across varying satellite altitudes, \\number of users, and task data sizes.} \\
        \hline
    \end{tabular}%
    }
\end{table*}
A hierarchical MEC system that integrates HAPS and UAVs was proposed in \cite{10506092}. The authors formulated a task offloading problem that minimizes the long-term cost of processing tasks while considering  queuing during task offloading. To solve it, they proposed a multi-agent DRL-based computation algorithm, enhanced by a convolutional LSTM (ConvLSTM) for UAVs' task load estimation and a prioritized experience replay (PER) method to accelerate the model's convergence and training stability.
In \cite{10373079}, the authors studied the duality of task offloading and computing resource allocation in a MEC-based aerial system of UAVs and HAPS, aiming to reduce energy consumption of ground devices. Specifically, they used stochastic optimization methods to split the problem into local resource computation sub-problems and solve them using a convex method, while for offloading resource allocation, they opted for a distributed game theoretical multiserver selection game (DGMS) algorithm and a transmission power allocation (TPA) algorithm. The aggregation of these methods, called DOTORA, achieved the least energy consumption, with reductions between $15\%$ and $19\%$, compared to conventional methods.
Finally, Ding \textit{et al.} studied in \cite{9515574} the integration of HAPS into LEO satellite networks in a satellite-aerial integrated edge computing network (SAIECN) framework, to minimize the system’s weighted sum energy consumption. They formulated the problem of HAPS-ground user association, multi-user multiple input and multiple output (MU-MIMO) precoding, computation task assignment, and resource allocation. To address it, they applied quadratic transform-based fractional programming (QTFP) and the weighted minimum mean-squared error method (WMMSE) to solve user association and MU-MIMO transmit precoding subproblems, respectively. The computation task assignment was solved using the classic interior point method, while the computation resource allocation was derived in a closed-form. 
Table \ref{tab:haps} provides a summary of the above works.

\subsection{UAV Network-based Computing}
UAV is also identified as one of the key technologies of B5G and 6G systems, given its capacity to improve communication and computing aspects of the network. For instance, UAVs acting as BSs, a.k.a., UAV-BSs, can enhance coverage and communication quality of cellular networks. Also, MEC-enabled UAVs can assist users lacking the necessary power to process data. 
For instance, in \cite{9725258}, the authors developed a collaborative MEC system with multiple UAVs and edge clouds to enhance service delivery for sparsely distributed users and reducing total system cost defined by the total execution delay and energy consumption. To do so, they jointly optimized UAVs trajectories, task allocation, and communication resources, using a cooperative multi-agent DRL (MADRL) framework with the twin delayed DDPG (TD3) algorithm. 
Authors of \cite{9897000} optimized the UAV's trajectory and task offloading decisions in a single UAV-assisted MEC system. They simultaneously minimized average task delay (ATD) and average energy consumption (AEC) while maximizing the average task number (ATN) collected by the UAV, using an improved evolutionary multi-objective RL method for trajectory control and task offloading, called EMORL-TCTO. 
Furthermore, Apostolopoulos \textit{et al.} investigated task offloading in a multi-UAV MEC-based network \cite{9392259}. They aimed to minimize the task offloading failure probability under latency and energy constraints. 
To solve the formulated problem, they modelled at first the UAV offloading decisions with Prospect theory, then used a DCP algorithm to obtain the pure Nash equilibrium (PNE) for the resource allocation non-cooperative game. The proposed approach converged to PNE in a few iterations, while achieving a low failure probability.
In \cite{9169676}, the authors proposed a collaborative UAV computing offloading scheme for IoT networks to maximize long-term network utility. They formulated the problem as a semi-MDP and solved it using a combined DQN and transfer learning. 
Xu \textit{et al.} introduced in \cite{9452794} the alternative computation efficiency maximization (ACEM) algorithm to optimize task offloading, computing, and communication resource allocation in UAV-assisted MEC-based networks. ACEM combined Dinkelbach’s method to transform the formulated fractional programming into parametric problems, Lagrange duality to achieve global optimal solutions for task offloading, CPU frequency selection, and bandwidth allocation, and SCA for UAV trajectory and time slot partition planning. 
In \cite{10540308}, the authors studied a MEC-based UAV-assisted network where UAVs act as MEC servers and blockchain nodes at the same time. To minimize the data processing delay, the authors proposed a block coordinate descent (BCD) iterative method that jointly optimizes UAVs' locations, data offloading, and resource allocation.
\begin{table*}[ht]
    \centering
    \huge
    \caption{Related Work on Computing in UAV Systems}
    \label{tab:uav}
    \resizebox{\textwidth}{!}{%
    \begin{tabular}{|c|c|c|l|l|}
        \hline
        \textbf{Ref} & \textbf{System Model} & \textbf{Objective} & \textbf{Proposed Solutions} & \textbf{Findings} \\
        \hline
        \cite{9725258} & \makecell{UAVs \\+ Edge clouds \\ + Users} & \makecell{Min. \\total execution delay \\ and energy consumption} & \makecell[l]{Cooperative MADRL for joint trajectory\\ planning, and task and power allocation.} & \makecell[l]{MADRL successfully optimizes task  offloading, \\ with the lowest cost compared to benchmarks.
        } \\
        \hline
        \cite{9897000} & \makecell{One UAV \\+ MEC-equipped BSs\\ + Smart devices} & \makecell{Min. task ATD and AEC, \\Max. ATN} & \makecell[l]{
        EMORL-TCTO algorithm for joint \\trajectory control and task offloading.} & \makecell[l]{
        EMORL-TCTO achieves a better overall balance \\ among the objectives, consistently outperforming\\ baselines.} \\
        \hline
        \cite{9392259} & \makecell{MEC-equipped UAVs \\ + MEC servers + Users} & \makecell{Min. task offloading\\ failure probability} & \makecell[l]{Prospect theory for UAV offloading decisions,\\ and  DCP to achieve PNE for resource allocation.}
        & \makecell[l]{
        Framework achieves the lowest failure probability\\ and converges fast to PNE compared to benchmarks.
        } \\
        \hline
        \cite{9169676} & \makecell{MEC-equipped UAVs \\ + MEC-equipped BSs \\ + IoT users} & \makecell{Max. total \\network utility} & \makecell[l]{DQN and transfer learning for joint cooperative \\offloading and resource allocation in \\centralized and distributed UAV networks.} & \makecell[l]{Proposed cooperative methods maximize long-term\\ network utility, outperforming non-cooperative \\approaches.} \\
        \hline
        \cite{9452794} & \makecell{UAVs \\+ MEC-equipped BSs \\ + Users} & \makecell{Max. computation \\efficiency} & \makecell[l]{ACEM algorithm integrates Dinkelbach’s method,\\ Lagrange duality, and SCA to jointly optimize \\task assignment and resource allocation.} & \makecell[l]{ACEM achieves the best performance by balancing \\computation bits and energy consumption \\for fixed numbers of users and MEC servers.} \\
        \hline
        \cite{10540308} & \makecell{Cloud computing \\ + MEC-equipped UAVs \\ + BSs + Users} & \makecell{Min. time consumption \\of data processing } & \makecell[l]{BCD-based algorithm to jointly optimize UAVs' \\locations, task offloading, and resource allocation.} & \makecell[l]{Compared to baselines, proposed method achieves \\ the lowest data processing time across various\\ settings, e.g., CPU frequency, data size, and area radii.} \\
        \hline
        \cite{10058597} & \makecell{SDN controller \\ + UAVs + MEC servers \\+ BSs + Users} & \makecell{Min. energy consumption \\and computation time} & \makecell[l]{OCRA with layers for resource allocation priority, \\resource optimization, and queue-based offloading.} & \makecell[l]{OCRA achieves the lowest energy consumption and \\computation time compared to benchmarks, in \\several conditions.
        } \\
        \hline
        \cite{10374285} & \makecell{MEC-equipped UAVs \\ + Users} & \makecell{Max. number of \\ offloaded tasks} & \makecell[l]{Two-layer DECO: Differential evolution\\ for UAV deployment, and convex optimization \\ for resource allocation.
        }
        & \makecell[l]{DECO algorithm achieves the highest performance\\ compared to baselines for different numbers of UAVs,\\ and delay constraints.} \\
        \hline
        \cite{10329934} & \makecell{MEC-equipped UAVs \\ + MEC servers + Users} & \makecell{Min. energy consumption \\of users and UAVs} & \makecell[l]{Greedy policy-based task offloading, \\ convex optimization for resource  allocation,\\ and SCA-based UAV location optimization.
        } 
        & \makecell[l]{
        Proposed framework achieves up to a 10.75\% \\reduction in energy consumption compared\\ to benchmarks.} \\
        \hline
        \cite{10480673} & \makecell{UAVs\\ + MEC-equipped RSUs\\ and Vehicles + Vehicle users} & \makecell{Min. service delay and \\ energy consumption} & \makecell[l]{GBATO: GBA for joint resource allocation\\ and offloading decision, and SCA-based method\\ for UAV trajectory optimization.
        } & \makecell[l]{
        GBATO realizes low delay and energy consumption \\performances, compared to other methods.}\\
        \hline
    \end{tabular}%
    }
\end{table*}
Goudarzi \textit{et al.} introduced in \cite{10058597} a two-layer cooperative evolutionary optimized computing resource allocation (OCRA) approach to minimize energy consumption and computation time in a UAV-assisted MEC network. The first layer prioritizes resource allocation, while the second optimizes resource utilization, supplemented by a queue-based offloading algorithm to determine optimal offloading decisions between ground MEC servers and UAVs. 
Similarly, the authors of \cite{10374285} introduced the differential evolution-based convex optimization (DECO) method to address the formulated MINLP problem, consisting of maximizing successful task offloading in a MEC-equipped UAV network. DECO is structured as a two-layer optimization approach: The upper layer employs a differential evolution algorithm to optimize 3D UAV deployment and elevation angles, while the lower layer uses convex optimization to make efficient task offloading and resource allocation decisions. 
In \cite{10329934}, the authors minimized the energy consumption of both ground devices and UAVs by jointly optimizing task offloading, transmission power, computing resource allocation, and UAV deployment. They opted for BCD to decouple the optimization problem into sub-problems, where task offloading is handled by a greedy-based algorithm, resource allocation by a convex optimization method, and UAV deployment by an SCA-based approach. 
Finally, Wang \textit{et al.} proposed in \cite{10480673} a cooperative computing framework for UAV-assisted vehicular networks, aiming to minimize service delay and energy consumption. In their system, a vehicle can also provide computing resources to other vehicles through V2V links, as shown in Fig. \ref{fig:schemaUAV}. The formulated problem has been decomposed into sub-problems using the Lyapunov framework. Then, a greedy bat algorithm (GBA) has been used for resource allocation and offloading decision, while SCA for UAV trajectory planning. The proposed approach that minimizes only the delay (resp. energy) is called GDOM (resp. GEOM), while the one that balances between delay and energy minimization is called GBATO.
In Table \ref{tab:uav}, we summarize the above discussed papers. 

\Figure[t!](topskip=0pt, botskip=0pt, midskip=0pt)[width=0.85\columnwidth]{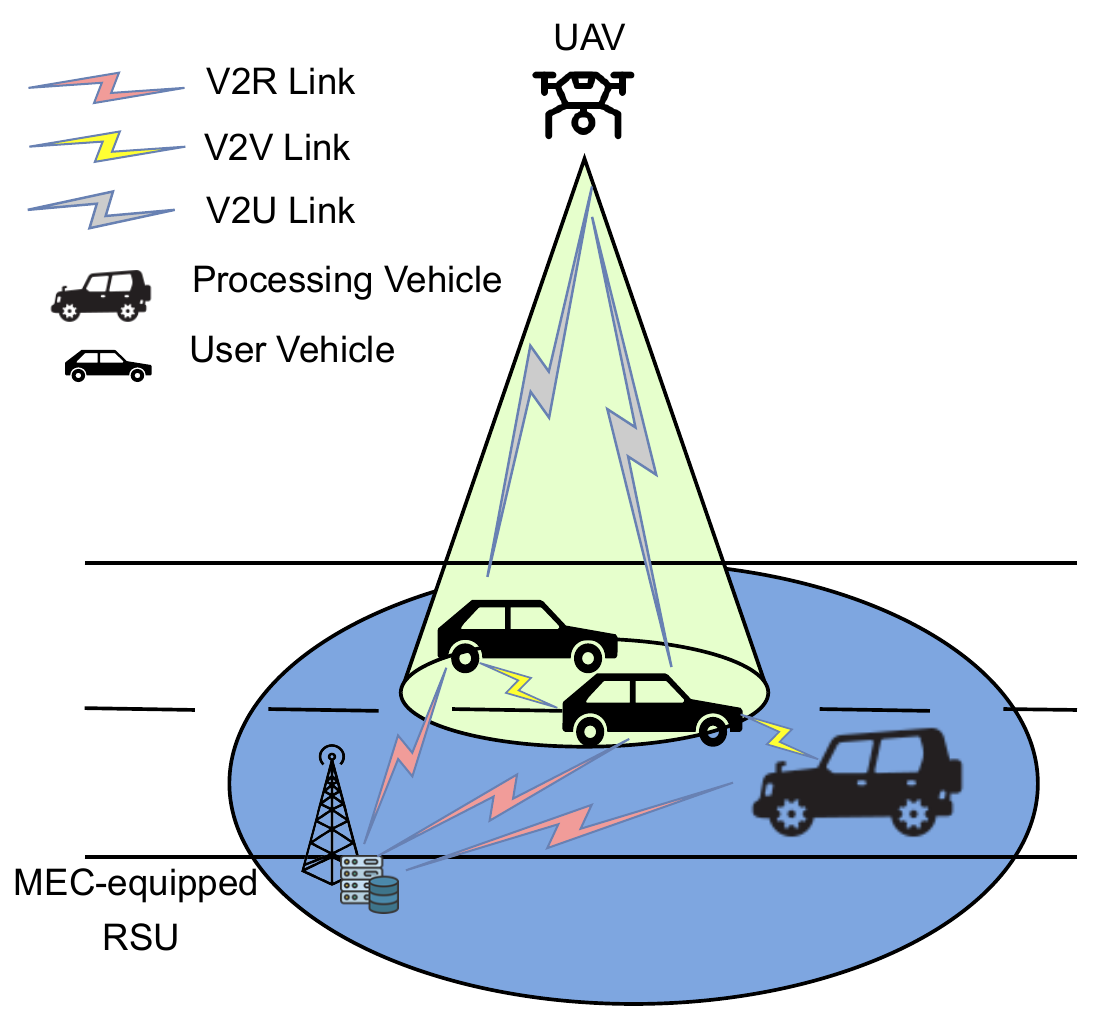}
{ \textbf{UAV-collaborative vehicular system model proposed \cite{10480673}.}\label{fig:schemaUAV}}

\section{ITNTN-based Computing}\label{sec:itntn}

The research on edge computing in all sections of TN and NTN has proven its value in contributing to a ubiquitous and integrated network. An edge-computing-based ITNTN can thus harness the full computing capacity of all layers and achieve a global connected network. This can be perceived as a response to the increasingly demanding applications for computing capacity and data. This section incorporates the research published on both partial and full ITNTNs. These works are summarized in Table \ref{tab:itntn} below.


The authors in \cite{9772280} proposed a three-layer computing framework comprising HAPS, RSUs, and vehicles where tasks are offloaded to minimize the total offloading delay. The multi-slot mixed-integer and continuous variable problems of task offloading, caching problems, and computing resource allocation were solved using a MARL method and Lagrangian-based allocation policy. Compared to the two-layer framework (without HAPS), the proposed solution achieved the lowest delay, with HAPS dominating the computing ratio at 70\%. 
A UAV and vehicular MEC-assisted computing framework is introduced to maximize energy efficiency in \cite{9964066}. The joint optimization problem of task offloading, computing resource allocation, and UAV trajectory is formulated to maximize energy efficiency. The latter is solved by a Lyanupov-based online algorithm.
The proposed approach minimized energy efficiency while maintaining shorter task queues compared to local and edge computing benchmark schemes.
Song \textit{et al.} designed in \cite{10147253} a similar computing framework for emergency services. Specifically, MEC-equipped UAVs and emergency vehicles support user tasks when BSs are saturated or unavailable.
They aimed to maximize the transmission rate and minimize delay and energy consumption. A horizontal federated learning - DDQN, a.k.a., HFL-DDQN, was proposed to solve the formulated multi-objective problem. 

In \cite{9793590}, the authors proposed a partially integrated computing network with LEO satellites and HAPS for vehicular task offloading. They aimed to minimize the total overhead of vehicular computation and communication by optimizing task offloading and computing resource allocation. For that, they proposed a decentralized value-iteration-based DDQL approach. 
Also, a computing-equipped space-air-ground integrated network (SAGIN) is introduced in \cite{10243118} to minimize the long-term completion delay. The formulated optimization problem is first decomposed into deterministic sub-problems with a Lyapunov method, then a delayed online learning technique is exploited to predict task arrival and queue length. Based on the latter, task offloading strategies can be adjusted dynamically. The proposed method, called online offloading and scheduling (OOS), achieved significant performance in terms of completion time and data rate for varying numbers of satellites, HAPS, and UAVs.
A similar SAGIN computing framework is introduced in \cite{9928786} to maximize the average sum rate of energy harvesting-artificial intelligence of everything (EH-AIoE) users. 
A Lyapunov-guided multi-agent proximal policy optimization (LMAPPO) was introduced to jointly optimize task offloading, computing resource allocation, and energy harvesting decisions. 
Moreover, the authors of \cite{10345669} integrated digital twin (DT) technology in their SAGIN computing framework for vehicular task offloading. 
To minimize the overall latency of the system, they proposed a MADRL approach that optimizes task offloading, energy consumption, and network coverage. The proposed method consumed $36.63\%$ less energy than the terrestrial MEC framework, at the expense of a slight increase in latency ($4.71\%$ more).
Authors in \cite{10138117} studied a cloud-satellite-UAV computing framework for natural disaster-struck areas. Their objective is to maximize the energy efficiency with respect to the QoE requirements. 
Dinkelbach and BCD methods were utilized to transform the objective function into a tractable form and to decouple the non-convex and non-concave optimization problem into sub-problems, which are solved iteratively. 
Huang \textit{et al.} proposed in \cite{10440193} a hybrid-MEC satellite-UAV computing system. To minimize latency and energy consumption, they introduced a DRL-based decision-assisted hybrid multi-agent soft actor-critic (DM-SAC-H) algorithm for joint task offloading and computing resource allocation.
A similar system was considered in \cite{10342725}, targeting to reduce energy consumption in UAVs and ground devices, while satisfying computing delay constraints. To solve the formulated problem, an AO-based method was adopted for user scheduling and partial offloading, and an SCA method for computing resource allocation, UAV trajectory, and bandwidth allocation. 
Shen \textit{et al.} introduced in \cite{10145842} a slicing-based collaborative offloading for an ITNTN framework supporting vehicular users. The goal is to maximize task completion rate under the constraint of coupled resources. An adaptive slicing window strategy was developed for offloading, which adapts its length according to traffic changes. Also, multidimensional resource orchestration and DDQN-based task scheduling were introduced. 

In \cite{9328513}, a satellite computing framework for aerial, ground, vehicular, and water surface users was proposed, where LEO satellites are equipped with computing power and GEO/MEO satellites act as communication relays. The objective is to reduce task completion time and satellite resource usage, using an offline deep imitation learning (DIL)-based offloading algorithm and a proactive-centralized caching method, a.k.a., DOCS approach. 
By leveraging SDN and NFV in \cite{9375493}, Cao \textit{et al.} designed a SAGIN framework for vehicular task offloading. They aimed to minimize multiple objective functions, including service delay, system energy consumption, resource utilization, and service security. To do so, they proposed the Two\_Arch2 algorithm and improved it with an angle-based diversity selection strategy. 
When compared against the NGSA-III and CA-MOEA benchmarks, the proposed method achieves the lowest delay and failure rates, and the highest utilization rate.

\begin{table*}[ht]
    \centering
    \huge
    \caption{Related Work on Computing in ITNTN Systems}
    \label{tab:itntn}
    \resizebox{\textwidth}{!}{%
    \begin{tabular}{|c|c|c|l|l|}
        \hline
        \textbf{Ref} & \textbf{System Model} & \textbf{Objective} & \textbf{Proposed Solutions} & \textbf{Findings} \\
        \hline
        \cite{9772280} & \makecell{MEC-equipped HAPS, \\RSUs, and Vehicles \\+ Vehicle users} & \makecell{Min. total \\offloading delay} & \makecell[l]{MARL method for joint task assignment \\ and caching decisions, and Lagrangian-based \\for computing resource allocation.
        } & \makecell[l]{Proposed approach achieves the lowest \\ offloading delay compared to two-layer\\ framework (without HAPS).\\ 
        } \\
        \hline
        \cite{9964066} & \makecell{MEC-equipped UAVs\\ and Vehicles + BSs \\ + Vehicle users} & \makecell{Max. energy\\ efficiency} & \makecell[l]{Lyapunov-based online optimization\\ for resource allocation and UAVs trajectory\\ planning.} & \makecell[l]{A trade-off between energy efficiency and\\ delay is realized, while maintaining low\\ queue lengths.} \\
        \hline
        \cite{10147253} & \makecell{MEC-equipped UAVs,
        \\Vehicles, and IoT\\
        + Users} & \makecell{Max. transmission rate \\
        Min. delay\\ and energy consumption} & \makecell[l]{HFL-DDQN for joint task offloading \\and relaying decisions.
        } & \makecell[l]{HFL-DDQN has the best performances \\ compared to benchmarks, in \\ particular for computation-intensive services.} \\
        \hline
        \cite{9793590} & \makecell{MEC-equipped LEO \\Sats. and HAPS \\+ Vehicle users} & \makecell{Min.
        overhead\\ of computation \\and communication} & \makecell[l]{Decentralized value-iteration-based DDQL
        \\for joint computation offloading \\and resource allocation.} & \makecell[l]{Proposed DDQL converges fast and \\ achieves low overhead for different MEC\\  capacity, task size, and complexity values.} \\
        \hline
        \cite{10243118} & \makecell{Cloud computing \\ + MEC-equipped LEO \\Sats., HAPS, and UAVs\\ + Users} & \makecell{Min. task \\completion delay} & \makecell[l]{OOS: Delayed online learning for task\\ arrival and workload prediction, and \\ collaborative online task scheduling policy.} & \makecell[l]{OOS realizes the lowest completion delay\\  and highest data rate compared to\\  benchmarks.} \\
        \hline
        \cite{9928786} & \makecell{MEC-equipped LEO\\ Sats., HAPS, and BSs \\ + EH-AIoE users} & \makecell{Max. avg. \\sum-rate\\ of EH-AIoE} & \makecell[l]{LMAPPO for joint optimization of task\\ offloading, computing resource allocation,\\ and energy harvesting decisions.
        } & \makecell[l]{LMAPPO has the best performance \\compared to baselines for varying\\ EH and computing conditions.} \\
        \hline
        \cite{10345669} & \makecell{MEC-equipped LEO \\Sats. and BSs + UAV\\  relays + Vehicle users} & \makecell{Min. overall\\ latency} & \makecell[l]{Proposition of a two-layer DT framework and \\of a MADRL algorithm to jointly optimize task \\offloading, energy consumption, and coverage.} & \makecell[l]{
        Proposed approach consumes less energy\\ than the terrestrial MEC framework, but\\  with a small increase in latency.} \\
        \hline
        \cite{10138117} & \makecell{Cloud computing \\+ MEC-equipped UAVs \\ + LEO Sat. + UAV relays \\+ Users} & Max. energy efficiency & \makecell[l]{
        Dinkelbach and BCD-based iterative algorithm\\ to jointly optimize UAV trajectory and\\  resource allocation.} & \makecell[l]{Compared to benchmarks, the proposed \\ method realizes near-optimal energy \\ efficiency for any network and task sizes.} \\
        \hline
        \cite{10440193} & \makecell{Cloud computing \\+ MEC-equipped LEO  \\Sats. and UAVs + Users} & \makecell{Min. avg. latency\\ and energy consumption} & \makecell[l]{DM-SAC-H: MADRL-based task offloading\\ and computing resource allocation.} & \makecell[l]{DM-SAC-H has the lowest energy\\ consumption compared to baselines\\ in different network conditions.
        } \\
        \hline
        \cite{10342725} & \makecell{Cloud computing\\ + MEC-equipped LEO \\Sats and UAVs + Users} & \makecell{Min. energy\\ consumption} & \makecell[l]{ 
        AO-based method for user scheduling and\\ partial offloading, and SCA for computing \\resource allocation, UAV trajectory  \\ planning, and bandwidth allocation.} & \makecell[l]{Proposed algorithm reduces weighted \\ sum energy consumption by up to\\ 36\% compared to benchmarks. 
        } \\
        \hline
        \cite{10145842} & \makecell{MEC-equipped LEO \\Sats., UAVs, and BSs \\+ Vehicle users} & \makecell{Max. task \\completion rate} & \makecell[l]{Collaborative task offloading with adaptive\\  slicing window strategy, Lagrange \\multiplier-based RAN resource allocation,\\ and DDQN-based task scheduling.} & \makecell[l]{Proposed framework achieves the\\ highest completion rate and lowest\\ failure rate compared to baselines for\\  different workloads, resources, etc.} \\
        \hline
        \cite{9328513} & \makecell{MEC-equipped LEO\\ Sats. + GEO/MEO Sat. relays\\ + Ground/Aerial/Marine users} & \makecell{Min. task completion time \\ and Sat. resource usage} & \makecell[l]{DOCS: DIL-based task offloading\\ and proactive-centralized caching.} & \makecell[l]{DOCS has the lowest completion time\\ and highest decision accuracy, compared\\  to benchmarks without learning.} \\
        \hline
        \cite{9375493} & \makecell{SDN/NFV network of\\
        Cloud computing + Sats. \\+ HAPS + UAVs + RSUs \\ + ground MEC servers \\+ Vehicule users} & \makecell{
        Min. service delay,\\ energy consumption,\\ resource utilization,\\ and service security} & \makecell[l]{SDN/NFV-based framework, in which\\ Two\_Arch2 algorithm is used for joint\\ optimization of task offloading\\ and computing resource allocation.} & \makecell[l]{Proposed algorithm achieves \\the best performances, compared to \\NGSA-III and CA-MOEA benchmarks.} \\
        \hline
    \end{tabular}%
    }
\end{table*}

\section{Open Issues and Future Research Directions}\label{sec:future}


In this section, we identify challenges and issues that require further investigation for the realization of computing-assisted ITNTN and future research directions. 

Integration of various technologies from different ITNTN layers, e.g., different protocols, propagation conditions, etc., incurs non-negligible delays during system convergence. Although research has shown promising convergence with interoperability, continuous improvement is essential. For instance, hybrid management can be leveraged where a centralized controller oversees the overall network and distributed controllers manage local adjustments. To make such control efficient, ML and FL techniques can be integrated, such that networks managed by different entities are able to benefit from each other's experience. 
To enhance ITNTN prediction and optimization capabilities, generative AI can be utilized to provide near-reality datasets. The latter would be used to enhance the management algorithms' accuracy and precision, without relying solely on the scarcely collected data from the real environment. Furthermore, large language models (LLMs), such as GPT-4, BERT, and LLaMA, can be used for the development of DT frameworks tailored to ITNTN systems and for decision explainability.



Moreover, energy efficiency has been a primary optimization focus in most of the studied works. Nonetheless, many ways to achieve green computing and networking are yet to be designed. For example, EH can leverage more efficient materials, including thermoelectric, solar, and piezoelectric technologies, to enhance EH performance. Also, ML can be exploited to build EH prediction systems and realize autonomous energy management. 
ITNTN can also benefit from the development of novel 6G concepts, such as Open-RAN (O-RAN) and reconfigurable intelligent surfaces (RIS). The former aims to democratize interfaces between software-based components to facilitate merging between TN and NTN components \cite{10439167}, while the latter improves SNR/SINR, extends coverage, and reduces interference \cite{10186454,Alfatt2021}.

Regulatory and frequency spectrum management can be challenging across different jurisdictions with high interference among the ITNTN layers' operation. To overcome the problem of spectrum attribution, the usage of higher frequency bands, such as millimeter-wave, can be leveraged, however further research is required for its application in NTN layers. Harmonization efforts of the protocols and standards of ITNTN must also be a priority to ensure its development.

Due to the distributed nature of ITNTN, the cyberattack surface is large and uncontrollable. 
In addition, in NTN layers, data interception and eavesdropping are inevitable due to the open nature of wireless communications. Thus, the development of robust security systems, based on strong encryption or frequency hopping techniques, is needed. Such efforts can be supported by the recent development of free-space optics (FSO), quantum communications, and quantum computing. Indeed, recent research has shown that quantum computing surpasses classical computing and can solve complex, specific problems at lower complexity \cite{10423012}.
Thus, it can be applied in ITNTN for reinforcing security through quantum cryptography, quantum-based optimization, etc.

Finally, the profitability of the ITNTN services will always be the driving force of its development, yet its economic model represents another intrinsic challenge. The cost of building a fully integrated system is high, while the integration of existing network layers would require adjustment, partly to create communication interfaces and ensure compatibility. The pricing model is also difficult to settle since it is important to provide an affordable-to-all service, and satisfy the profit expectations of each ITNTN layer(s) operator. 

\section{Conclusion}\label{sec:end}
In this paper, we provided an overview of the current landscape of computing within ITNTN. After presenting the fundamentals of computing, we extensively reviewed the most recent works studying computing partially or fully within ITNTN layers. 
For each work, we exposed the system model, the tackled problem, and the techniques employed to solve it. 
In achieving their goals, these works called for solutions ranging from classical functional analysis to ML, highlighting their efficacy in comparison to others. Finally, we identified open issues and research directions pertaining to computing within ITNTN.



\vspace{-10pt}
\bibliographystyle{IEEEtran}
\bibliography{IEEEabrv,Biblio}

\EOD

\end{document}